\documentclass[twocolumn]{article}
\pdfoutput=1

\DeclareRobustCommand{\VAN}[3]{#2}
\let\VANthebibliography\thebibliography
\def\thebibliography{\DeclareRobustCommand{\VAN}[3]{##3}\VANthebibliography}

\usepackage{newtxtext,newtxmath}
\usepackage[T1]{fontenc}

\DeclareRobustCommand{\VAN}[3]{#2}
\let\VANthebibliography\thebibliography
\def\thebibliography{\DeclareRobustCommand{\VAN}[3]{##3}\VANthebibliography}

\usepackage{arxiv}
\usepackage{graphicx}	
\usepackage{amsmath}	
\usepackage{amsfonts}
\usepackage{mathrsfs}
\usepackage{subcaption}
\usepackage[retainorgcmds]{IEEEtrantools}
\usepackage{gensymb}
\usepackage{verbatim}
\usepackage{tikz}
\usepackage{tikz-3dplot}
\usepackage{siunitx}
\usepackage{color}
\usepackage{pdflscape}
\usepackage{commath}
\usepackage{xcolor}
\usepackage{natbib}
\usepackage[utf8]{inputenc} 
\usepackage{hyperref}       
\usepackage{url}            
\usepackage{booktabs}       
\usepackage{nicefrac}       
\usepackage{microtype}      
\usepackage{doi}

\DeclareSIUnit{\parsec}{pc}
\DeclareSIUnit{\electronvolt}{eV}
\DeclareSIUnit{\eV}{\electronvolt}
\DeclareSIUnit{\keV}{\kilo\electronvolt}

\usepackage{authblk}
\usepackage{caption}

\newcommand*{\mathcolor}{}
\def\mathcolor#1#{\mathcoloraux{#1}}
\newcommand*{\mathcoloraux}[3]{%
  \protect\leavevmode
  \begingroup
    \color#1{#2}#3%
  \endgroup
}

\title{\texttt{Skylight}: a new code for general-relativistic ray tracing and radiative transfer in arbitrary spacetimes}

\author[1,2]{Joaquín Pelle}
\author[1,2]{Oscar Reula}
\author[1,2,3]{Federico Carrasco}
\author[1,2]{Carlos Bederian}

\affil[1]{\textit{\normalsize{Facultad de Matem\'atica, Astronom\'ia, F\'isica y Computaci\'on, Universidad Nacional de C\'ordoba, Argentina}}}

\affil[2]{\textit{\normalsize{Instituto de F\'isica Enrique Gaviola, CONICET, Ciudad Universitaria, 5000 C\'ordoba, Argentina}}}

\affil[3]{\textit{\normalsize{Max Planck Institute for Gravitational Physics (Albert Einstein Institute), 14476 Potsdam, Germany}}}

\date{}

\renewcommand{\headeright}{}
\renewcommand{\undertitle}{A Preprint}
\renewcommand{\shorttitle}{\texttt{Skylight}: a new code for relativistic ray tracing and radiative transfer in arbitrary spacetimes}

\begin{document}

\twocolumn[
\begin{@twocolumnfalse}
\maketitle

\begin{abstract}

To reproduce the observed spectra and light curves originated in the neighborhood of compact objects requires accurate relativistic ray-tracing codes. In this work we present \texttt{Skylight}, a new numerical code for general-relativistic ray tracing and radiative transfer in arbitrary space-time geometries and coordinate systems. The code is capable of producing images, spectra and light curves from astrophysical models of compact objects as seen by distant observers. We incorporate two different schemes, namely Monte Carlo radiative transfer, integrating geodesics from the astrophysical region to distant observers, and camera techniques with backwards integration from the observer to the emission region. The code is validated by successfully passing several test cases, among them: thin accretion disks and neutron stars hot spot emission.  
\end{abstract}


\hfill
\hfill
\hfill
\hfill

\end{@twocolumnfalse}
]



\section{Introduction}



The theory of general relativity (GR) has been tested under a wide variety of circumstances, ranging from planetary to cosmological scales, and within different degrees of nonlinearity. The most elusive tests, still to be completed, remain those in the strong field regime. Compact relativistic objects, including neutron stars and black holes, are excellent natural laboratories where extreme phenomena takes place allowing us to probe the behavior of matter under the influence of very strong gravitational fields. In these scenarios, the gravitational field plays a crucial role in the astrophysical processes which occur, leaving an imprint on the observations by shifting the energies and deflecting the trajectories of the emitted photons.  In turn, this also allows us to put GR to the test at the very instances where its nonlinearities are the strongest.

For example, X-ray pulse profiles generated from hot spots on the surface of spinning neutron stars are severely affected by the gravitational field of the star \citep[][]{pechenick1983hot}. The emission lines produced in black hole accretion disks suffer relativistic broadening due to the combined effect of gravitational redshift and Doppler boosting to the fluid frame \citep[][]{miller2002evidence}.

In the last few years, new instruments with increased sensitivity like NICER \citep{gendreau2012neutron} began to operate producing very accurate observational data. The forthcoming mission eXTP \citep[][]{zhang2019enhanced} will also add up to this in the near future. The modelling of pulse profiles and comparison with these high-quality data can be used, e.g., to constrain the mass-radius relation of neutron stars, and consequently their equations of state and interior compositions \citep[][]{riley2019nicer, riley2021nicer,pang2021nuclear}. 
It has also been used to infer the possible topology of the magnetic field near the stellar surface, suggesting the existence of global-scale multipolar components in millisecond pulsars (particularly for PSR J0030+0451) \citep[][]{bilous2019nicer,chen2020numerical,kalapotharakos2021multipolar}. Thus, challenging the standard pulsar picture consisting of a centered magnetic dipole which would yield two antipodal emitting polar cap regions.

Recently, the Event Horizon Telescope collaboration obtained an image of the black hole M87\textsuperscript{*}, being the first image of a black hole ever captured \citep[][]{akiyama2019first}. This opened unprecedented possibilities to deepen our understanding of physics in these extreme regimes, such as investigating the gravitational fields and charges of black holes \citep[][]{psaltis2020gravitational, kocherlakota2021constraints}, the magnetic field structure near the event horizon \citep{akiyama2021first}, the jet launching and collimation mechanisms \citep{jeter2020differentiating}, and so on.

To theoretically reproduce the spectrum and light curves of these sources with high precision requires accurate general-relativistic ray-tracing and radiative transfer codes. A number of such codes have been developed for that purpose, generally adjusting to one of two schemes: e.g. \cite{schnittman2004harmonic, noble2007simulating, dexter2009fast, psaltis2011ray, bronzwaer2018raptor, pihajoki2018general} and \cite{moscibrodzka2018ipole}, which follow observer-to-emitter schemes, i.e. tracing rays from a virtual observer to the source backwards in time; and, on the other hand, \cite{dolence2009grmonty} and \cite{schnittman2013monte} that follow Monte Carlo schemes in which photon packet distributions are sampled at the source and are later propagated outwards. 

In this paper, we present a new numerical general-relativistic ray-tracing and radiative transfer code, \texttt{Skylight}, and we demonstrate its accuracy and appropriateness for astrophysical applications. \texttt{Skylight} supports transfer in arbitrary asymptotically-flat space-time geometries and coordinate systems. The reason we adopted this geometry-agnostic position for our code is that in the near future we will incorporate approximate and numerical metrics to investigate emission models in binary systems, systems for which no exact metrics are known. Both the observer-to-emitter and emitter-to-observer schemes are implemented in \texttt{Skylight}, a property only shared with the code described in \cite{schnittman2013monte}. While the observer-to-emitter scheme demands less allocations and computational time, often working fine in a laptop, the emitter-to-observer is more amenable to the inclusion of scattering processes in a future generalization. The code is capable of producing images, phase-resolved and phase-averaged spectra, light curves, sky maps and animations. The code may also be applied to any astrophysical problem involving radiation transport in a curved spacetime, not necessarily in the presence of compact objects, as in the propagation of light at cosmological scales. The ray-tracing facility may also be used to compute the trajectories of massive particles in an arbitrary spacetime.

Our ray-tracing algorithm is natively written in the relatively new high-performance dynamically-typed language Julia. In the past, \cite{McKinnon2015RelativisticRT} has ported to Julia the Python ray tracer STARLESS\footnote{https://github.com/rantonels/starless}, but it is restricted to the Schwarzschild spacetime. \texttt{Skylight} has the first Julia ray tracer that is able to handle arbitrary space-time geometries. This paper also serves as a demonstration of the suitability of Julia for scientific astrophysical problems.

One of our main goals in developing \texttt{Skylight} is to use it in combination with the 3D general-relativistic force-free code \texttt{Onion} \citep{carrasco2017novel}.
Many relevant astrophysical scenarios involving compact objects are likely to be filled by a magnetically dominated plasma, well suited to the force-free (FF) approximation. Such plasma environments would typically allow to channel a fraction of the available kinetic energy into the sourrounding electromagnetic field; energy which can be then reprocessed within the magnetosphere to produce emissions on the different bands of the electromagnetic spectrum. The main limitation of the FF approach is, however, that it does not directly account for particle acceleration and micro-physical processes responsible of producing the actual electromagnetic signals.
Hence, several strategies were developed --mainly in the study of pulsars-- to connect the global electromagnetic field configurations provided by the FF description with the micro-physics involved in the emission processes (e.g., \cite{bai2010, lockhart2019x, chen2020numerical, kalapotharakos2021multipolar}). 
The idea is to first numerically solve the magnetosphere of different relevant systems within the FF approximation \citep[see e.g.][]{carrasco2018pulsar, carrasco2019triggering, carrasco2021magnetospheres} and then use these solutions as a starting point to model possible electromagnetic emissions processes and compute their associated light curves and spectra with \texttt{Skylight}.

The structure of this paper is as follows: in Section~\ref{sec:physical_setup}, we present the basic physical setup of the ray-tracing and radiative transfer problem. In Section~\ref{sec:description_code}, we describe the code giving the details of both the emitter-to-observer and observer-to-emitter schemes, including the initial data setting, the ray-tracing and the flux calculation steps. We show validation tests of the ray-tracing integrator in Section~\ref{sec:verification}, using the constants of motion of the Kerr spacetime and comparing with a semi-analytic ray-tracing function in the Schwarzschild spacetime. In Section~\ref{sec:astrophysical_tests}, we validate the complete structure of radiative transfer within the context of some astrophysical test applications. Then, we present tests of numerical convergence for the observer-to-emitter and emitter-to-observer schemes in Section~\ref{sec:convergence}. Finally, we summarize our work and conclude in Section~\ref{sec:conclusions}.

\section{Physical setup}
\label{sec:physical_setup}

\subsection{Geodesic equations}

In the general theory of relativity (GR) spacetime is represented by a four-dimensional Lorentzian manifold. The metric $g_{\mu \nu}$ is a symmetric non-degenerate rank-2 tensor field over spacetime and represents the gravitational field. \texttt{Skylight} supports arbitrary space-time geometries and coordinate systems, provided only that they are asymptotically flat. The geometry enters the code simply via the components of the metric and the Christoffel symbols,
\begin{align}
    \Gamma^\alpha_{\mu \nu} = \frac{1}{2}g^{\alpha \rho}(\partial_\mu g_{\nu \rho} + \partial_\nu g_{\mu \rho} - \partial_\rho g_{\mu \nu})\,,
\end{align}
written as functions in the coordinate system of choice. 

In GR, freely falling test particles follow the timelike or null geodesics of the spacetime. The equations of the geodesics are
\begin{equation}
\begin{aligned}
    \frac{\mathop{d^2 x^{\alpha}}}{\mathop{d\lambda^2}}+\Gamma^\alpha_{\mu \nu} \frac{\mathop{d x^{\mu}}}{\mathop{d\lambda}} \frac{\mathop{d x^{\nu}}}{\mathop{d\lambda}}=0\,, 
\end{aligned}
\label{eq:geodesic}
\end{equation}
where $x^\alpha$ is the position of the particle and $\lambda$ is proper time in the timelike case and an affine parameter in the null case. The type of geodesic is determined by the mass of the particle, where massive particles follow timelike geodesics and massless particles follow null geodesics.

Photons are no exception to this principle of geodesic motion. Thus, gravity can deflect them and redshift their energies. Therefore, light curves and spectra will be severely affected by space-time curvature whenever the photons are emitted close to a strong gravitational field source like a black hole or a neutron star.

\subsection{Covariant transport}
\label{sec:transfer}
The radiation field on a curved spacetime can be covariantly described in terms of a Lorentz invariant phase-space photon density, $\mathcal{F}(x^{\mu},k^{\mu})$, where $x^{\mu}$ is space-time position and $k^{\mu}$ is four-momentum \citep[][]{lindquist1966relativistic},
\begin{equation}
    k^{\mu} = \frac{\mathop{dx}^{\mu}}{\mathop{d\lambda}}\,.
\end{equation}
The invariant density is related to the specific intensity of the radiation field via $\mathcal{F} = \nu^{-3} I_\nu$, where $\nu$ is the photon frequency. Note that $\nu^3$ and $I_\nu$ are not separately Lorentz invariant. We adopt this latter description in terms of specific intensity, which is more commonly used. In these terms, the covariant transport equation along a geodesic reads
\begin{align}
    \frac{\mathop{d}}{\mathop{d\lambda}} \left( \frac{I_\nu}{\nu^3}\right) &= \frac{j_\nu}{\nu^2} - \nu \alpha_\nu \left( \frac{I_\nu}{\nu^3}\right) \,,
    \label{eq:complete_transfer}
\end{align}
where $\lambda$ is the affine parameter of the geodesic and $j_\nu$ and $\alpha_\nu$ are the emissivity and absorptivity of the medium, respectively. Each term in the equation is Lorentz invariant. The operator $d / \mathop{d \lambda}$ is the Liouville operator, i.e. the convective derivative in phase space. However, once the geodesic is given, the operator acts just as an ordinary derivative with respect to $\lambda$. 

For the moment, we have not included cases with $\alpha_{\nu} \neq 0$, and in our current applications the support of $j_\nu$ is contained within a three-surface which is spatially compact. This includes, for example, the emission from neutron star hot spots and thin accretion disks. We have left the general case for a future work.

In vacuum, the density $\mathcal{F}(x^{\mu},k^{\mu})$ is also invariant under the geodesic flow, since
\begin{align}
    \frac{\mathop{d}}{\mathop{d\lambda}} \left( \frac{I_\nu}{\nu^3}\right) &= 0 \,,
    \label{eq:vacuum_transfer}
\end{align}
i.e. its value is constant along the geodesic generated by $k^{\mu}$ at the point $x^{\mu}$. Therefore, in vacuum, the task essentially consists in obtaining enough solutions to the geodesic equations with different starting points and momentum vectors and connecting the information between the extreme points. This is done by using the fact that $I_{\nu_0} / \nu_0^3 = I_\nu / \nu^3$, where $\nu_0$ is the frequency at the starting point and $\nu$ is the gravitationally redshifted frequency at the endpoint. 

\section{Code description}
\label{sec:description_code}

\texttt{Skylight} has two different schemes of operation. On the one hand, an emitter-to-observer scheme, in which the local emissivity is sampled as a distribution of photon packets, and photons are propagated up to a large distance where virtual observers are located. And on the other hand, an observer-to-emitter scheme, in which a virtual detector is set at the location of the observer, and from every pixel the path of a past-directed photon is traced towards the emitting source.

Most of our applications will correspond to the emission generated in a region which rotates stationarily around an axis, as is usually the case, e.g., in spinning neutron stars and black hole accretion disks. Under this circumstance, all physical quantities depend on time and azimuth only via the combination $\omega t - \varphi$, namely the angular phase. Here $\omega$ is the angular rotation frequency of the system. Throughout the description below, after dealing with the general case, we will emphasize this particular instance of stationary rotation, as it simplifies the treatment and we will use it often. 

The most computationally demanding part of the code is the ray tracing. The integration of equations~(\ref{eq:geodesic}) for millions of rays throughout large distances requires a high-performance programming language. This part of the code is written in the relatively new high-performance dynamically typed language Julia. In particular, we use the package DifferentialEquations.jl \citep[][]{rackauckas2017differentialequations}. In this sense, \texttt{Skylight} also serves as a proof of the suitability of Julia and the mentioned package for high-performance scientific computing in astrophysics. 

The setting of the initial data and the post-processing of the output data are written in Python, and the whole code is integrated via a metadata management structure.

In Section~\ref{sec:ray_tracing}, we describe our ray-tracing algorithm. Later, in Sections~\ref{sec:direct} and \ref{sec:reverse}, we give the details of the initial data and post-processing in both transport schemes of the code.

\subsection{Ray-tracing algorithm}
\label{sec:ray_tracing}

Instead of integrating the equations~(\ref{eq:geodesic}) in its second-order form directly, \texttt{Skylight} integrates its enlarged first-order form: 
\begin{equation}
\begin{aligned}
    \frac{\mathop{dx^{\alpha}}}{\mathop{d\lambda}} &= k^\alpha\,,\\
    \frac{\mathop{dk^{\alpha}}}{\mathop{d\lambda}} &= -\Gamma^\alpha_{\mu \nu} k^\mu k^\nu\,, 
\end{aligned}
\label{eq:geodesicfirst}
\end{equation}
where $k^\alpha$ is the four-momentum of the photon. This formulation avoids computational issues at radial turning points, where the signs of the first derivatives would have to be checked to proceed in the second-order formulation. More importantly, adopting this formulation allows us to use the standard methods for the numerical solution of first order systems of ODEs.

For the numerical integration of the differential equations we use the Julia package DifferentialEquations.jl \citep[][]{rackauckas2017differentialequations}, which provides a wide variety of built-in algorithms for the numerical solution of ODEs, much wider than traditional libraries. Apart from the standard algorithms, this library includes many algorithms which are the result of recent research and are known to be more efficient than the traditional choices. Thus, we have at our disposal many different methods to choose according to the size of the system, the required accuracy, the presence of stiffness, the need of adaptivity, the available storage, etc. The package also counts with an efficient automated solver selector \citep[][]{rackauckas2019confederated} and it easily admits parallelization. 

Even though we are not bound to any particular method, we are specially fond of the method VCABM \citep[][]{hairer1993solving}, an adaptive-order adaptive-time Adams-Moulton method, which is a good choice for high accuracy in very large systems as the ones we deal with. Step size adaptivity also comes in handy: small steps are required close to the source of the gravitational field in order to preserve accuracy, but far away from the source, geodesics are approximately straight lines, so large steps are convenient there to reduce computational costs. VCABM is the solver method we used in all the applications presented in this paper. The relative and absolute tolerances can be set as parameters of the method. See Section~\ref{sec:verification} for a validation of our ray-tracing algorithm.

The cutoff conditions for the integration of the geodesics depend on the scheme (emitter-to-observer or observer-to-emitter) and on the particular problem. For example, in the emitter-to-observer where photons are propagated outwards from the source, the integration of a geodesic would terminate when it arrived at a sufficiently large distance, where virtual detectors are supposed to be located. On the other hand, in the observer-to-emitter scheme, the geodesics would be integrated  from the image plane until they arrived upon the emitting surface or they strayed too far from the source without having intersected it. If, for example, there was a black hole in the numerical domain, the geodesic integration would terminate whenever it entered the event horizon, since thereupon it could never exit that region. 

\subsection{Emitter-to-observer scheme}
\label{sec:direct}

\subsubsection{Initial data}
\label{sec:dirinidat}

In this scheme, the emission model enters the code via an emissivity distribution, $j_\nu$, which depends on space-time position, frequency and direction of emission. This distribution encodes the relevant information about the astrophysical processes occurring in the region of interest. In terms of the emissivity, the photon number density satisfies
\begin{equation}
    dn = \frac{j_\nu}{h \nu} \sqrt{-g}  d^{4}x d\nu d\Omega\,,
    \label{eq:initial-distribution}
\end{equation}
where $h$ is the Planck constant, $\nu$ is the photon frequency, $d\Omega$ is the solid angle element and $\sqrt{-g} d^4x$ is the invariant volume element. The frequency and the direction of emission are referred to the frame where the emissivity is defined. In most cases there is a preferred class of frames, namely the orthonormal frames where the local phenomena giving origin to the photons are at rest, which we call local comoving frames. These frames are of the form $\{ e^\mu_{(a)}: 0 \leq a \leq 3 \}$, where the greek letter is a contravariant vector index and the latin letter is a label, satisfying
\begin{equation}
    g_{\mu \nu} e^\mu_{(a)} e^\nu_{(b)} = \eta_{(a)(b)}\,,
    \label{eq:orthonormality}
\end{equation} 
where the right-hand side is the flat metric in its diagonal form. For the emission region to be at rest in this frame, the timelike vector field must equal the four-velocity of the emitting material, i.e. $e^\mu_{(0)} = u^\mu$. For the rest of the vector fields ---the spacelike vector fields--- there is a certain degree of freedom, as long as they satisfy equation~(\ref{eq:orthonormality}). Whenever we require them, we calculate the spacelike vectors are obtained by orthonormalizing a trial set of spacelike vectors via a Gram-Schmidt algorithm.

For representing the local emissivity, we must sample a distribution of photon packets following
\begin{equation}
    dN = \frac{dn}{w} = \frac{1}{w}\frac{j_\nu}{h \nu} \sqrt{-g}  d^{4}x d\nu d\Omega\,,
    \label{eq:initial-distribution-packets}
\end{equation}
where $w$ is the weight of the packet, i.e. the relative amount of photons it carries. This resembles what is done in Monte Carlo simulations. Associating a weight to the packets is not strictly necessary at this instance, but might be very convenient in some situations as we will explain later.

In the first place, we take a set of initial space-time positions distributed according to the momentum-integrated version of equation~(\ref{eq:initial-distribution-packets}). In the case of a stationarily rotating system, the initial time can be taken as $t=0$ for all packets, deferring all timing considerations to the post-processing of the output data, as we describe in Section~\ref{sec:flux_direct}.

Then, at each initial point we do a random sampling of the initial four-momenta of the photon packets. The four-momentum of a packet can be written as 
\begin{equation}
k^{\mu}= k^{(a)} e^{\mu}_{(a)}\,,
\label{eq:tetrad_components}
\end{equation}
where $k^{(a)}=\nu(1,\mathbf{\Omega})$ are the momentum components in the local comoving frame, $\nu$ is the frequency and $\mathbf{\Omega}$ is the direction of emission. The frequency and angular distributions are sampled according to equation~(\ref{eq:initial-distribution-packets}) evaluated at each point. In the case where $\alpha_\nu=0$, a single frequency can be taken as a representative of the entire spectrum, avoiding spectral sampling, since the trajectories do not depend on frequency. 

Finally, we convert the four-momenta to the coordinate frame according to equation~(\ref{eq:tetrad_components}), using the frame vectors $e^{\mu}_{(a)}$ calculated at each initial point. Once the initial set of packets is ready, we propagate them as described in Section~\ref{sec:ray_tracing} up to a large distance where virtual detectors are located.

\subsubsection{Flux calculation}
\label{sec:flux_direct}
The virtual detectors are located at a distance large enough so that curvature is negligible there, and, hence, the analysis can be done as in Euclidean geometry (recall that our spacetime is required to be asymptotically flat). The effects of curvature, e.g. redshift and deflection of photon trajectories, have been already encoded in the map relating the initial and the final data sets. 

In practice, the virtual detectors are simply small bins on the celestial sphere. In order to measure the monochromatic flux through a detector at inclination $\xi$ and azimuth $\varphi$, we collect the photons passing through it, and bin the ranges of frequency and time. Then, we calculate the flux as 
\begin{equation}
    F_{\nu} = \frac{1}{D^2 \Delta \Omega \Delta t \Delta \nu } \sum_i (h \nu)_i w_i \,,
\end{equation}
where $D$ is the distance to the observer, $\Delta \Omega \approx \sin \xi \Delta \xi \Delta \varphi$ is the solid angle occupied by the detector, $\Delta t$ is the size of a small temporal bin, and $\Delta \nu$ the size of a small frequency bin. The sum is over all photon packets collected by the detector. In this manner, we can produce phase-resolved and phase-averaged spectra, and by integrating on spectral windows we can also obtain sky maps and light curves. By taking into account the final direction of the photon packets three-momenta we can also produce images of the emitting source.

In the stationarily rotating case, all physical quantities depend on time and azimuth only via the angular phase $\omega t - \varphi$. In particular, the flux also depends on time and azimuth only via $\omega t - \varphi$, so we only need to look at the flux corresponding to detectors at $\varphi=0$ for different inclinations. We mentioned before that this symmetry allows us to sample and evolve a single set of initial photon packets departing from the source at the same time coordinate $t$. With the final positions of these photon packets in the celestial sphere and taking advantage of the symmetry of our system we can calculate everything we need. If we want to know the flux corresponding to a detector at an inclination $\xi$ and azimuth $\varphi=0$, we can concentrate on all the photon packets of our final data set which satisfy $|\xi-\xi_\mathrm{f}| \leq \Delta \xi /2$ and $0 \leq \varphi_\mathrm{f} < 0$, where $\xi_\mathrm{f}$ is the final polar angle of the photon and $\varphi_\mathrm{f}$ its final azimuth. This region is an annular strip of width $\Delta \xi$ on the celestial sphere centered at the inclination $\xi$ of the detector, and occupying a solid angle $\Delta \Omega \approx 2\pi \sin \xi \Delta \xi$. Let us suppose a photon (which departed at $t=0$ from the source) arrives to this annular strip at a time $t_{\mathrm{f}}$ with a final azimuth $\varphi_\mathrm{f}$ (not necessarily at $\varphi_\mathrm{f}=0$, where the detector lies). Then, due to the symmetry of our system, another photon (possibly emitted at a different initial time) would arrive at the detector with the same properties as the former, but at a different final time. The time of arrival of this latter photon to the detector can be computed as follows. For convenience, let us first define the observation phase as $\phi = (\omega t - \varphi)/2\pi$; at the location of the detector, the relation between time and observation phase reduces to $t = \phi T$, where $T = 2\pi / \omega$ is the period of the system. Thus, the latter photon will arrive at the detector with an observation phase
\begin{align}
 \phi &=-\frac{\varphi_\mathrm{f}}{2\pi}+\frac{t_\mathrm{f}}{T}\,.   \label{eq:time-delay}
\end{align}
Therefore, by the formula above we can assign an observation phase to all the photon packets which arrive to the annular strip. Then, in terms of the observation phase, the monochromatic flux at the detector can then be calculated as
\begin{equation}
    F_{\nu} = \frac{1}{ 2\pi D^2 \sin \xi \Delta \xi T \Delta \phi \Delta \nu } \sum_i (h \nu)_i w_i \,,
\end{equation}
where we have used that $\Delta t = T \Delta \phi$ at the detector, and the sum is over all photon packets which lie in the annular strip, within the corresponding phase and frequency bins.

\subsection{Observer-to-emitter scheme}
\label{sec:reverse}

Recall that spacetime is required to be asymptotically flat, therefore at distances as large as those of the observers, the analysis can be carried out as in flat spacetime. Let us take an inertial coordinate system $(t,x,y,z)$ such that the observer is at rest on the $xz$-plane at a distance $D$ from the origin, and at an inclination $\xi$ with respect to the $z$-axis. The monochromatic flux through a surface element at the location of the observer (the detector) is related to the specific intensity of the radiation field via
\begin{equation}
    F_\nu(t) = \int_{\mathcal{U}} I_\nu(\mathbf{\Omega},t) \mathbf{n} \cdot \mathbf{d\Omega} \,,
    \label{eq:flux_camera_angle}
\end{equation}
where $\mathcal{U}$ is a solid angle containing the source, $\mathbf{n}$ is the three-vector normal to the surface element, and all quantities are evaluated at the position of the observer. 

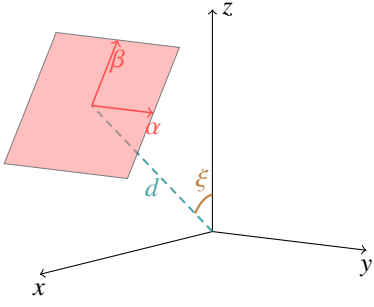
\begin{figure}
    \centering
   
    \tdplotsetmaincoords{80}{125}
    \begin{tikzpicture}[tdplot_main_coords]
    \coordinate (O) at (0,0,0);
    
    \draw[->] (O) --++ (4,0,0) node[below] {$x$};
    \draw[->] (O) --++ (0,2.5,0) node[below] {$y$};
    \draw[->] (O) --++ (0,0,3) node[right] {$z$};

    \coordinate (P) at  (2.795224285165525, 0.0, 2.1063525810321693) ;

    \draw[teal,thick,densely dashed,opacity=0.7] (O) -- (P) node[midway,below] {$d$};
    
    \tdplotdefinepoints(0,0,0)(0,0,0.5)(0.39931775502364647, 0.0, 0.3009075115760242);
    \tdplotdrawpolytopearc[brown,thick]{0.5}{anchor=south,color=brown}{$\xi$};

    \filldraw[fill=red!50,opacity=0.5]  (3.3970393083175736, -1.0, 1.3077170709848764)  --  (3.3970393083175736, 1.0, 1.3077170709848764)  --  (2.1934092620134766, 1.0, 2.904988091079462)  --  (2.1934092620134766, -1.0, 2.904988091079462)  --  cycle;

    \draw[semithick,red!70,->] (P) --  (2.795224285165525, 1.0, 2.1063525810321693)  node[below] {$\alpha$};
    \draw[semithick,red!70,->] (P) --  (2.1934092620134766, 0.0, 2.904988091079462)  node[below] {$\beta$};

    \end{tikzpicture}
    \caption{Sketch of the initial data setting in the observer-to-emitter scheme, showing the rectangular coordinates $(\alpha,\beta)$ on the image plane, the inclination of the observer $\xi$ and the distance to the image plane $d$.}
    \label{fig:reverse-scheme}
\end{figure}

Since the detector is far away from the source, the light rays arriving to it are almost exactly parallel. Thus, considering that for such rays $\mathbf{n} \cdot \mathbf{d\Omega} \approx d\Omega$ and using $d\Omega = dA / D^2$, we can rewrite the integral of equation~(\ref{eq:flux_camera_angle}) in the following form:
\begin{equation}
    F_\nu(t) = \frac{1}{D^2} \int_{\mathcal{S}} I_\nu(\alpha,\beta,t) d\alpha d\beta \,,
    \label{eq:flux_camera}
\end{equation}
where $\mathcal{S}$ is an image plane perpendicular to the line of sight, and $(\alpha,\beta)$ are rectangular coordinates over $\mathcal{S}$. These coordinates are related to the inertial coordinates via
\begin{align}
    x &= -\beta \cos \xi + d \sin \xi\,, \\
    y &= \alpha\,, \\
    z &= \beta \sin \xi + d \cos \xi \,,
\end{align}
where $d$ is the distance of the virtual detector. The setting is shown in Fig.~\ref{fig:reverse-scheme}. Notice that $\mathcal{S}$ need not be located at the true distance $D$, but a smaller distance $d$ is acceptable provided the effects of gravity in the direction of travel are negligible as well, since in such a case the result of the integral depends only on the impact parameters of the light rays and not on $d$ itself. For this reason, although our real astrophysical sources of interest might be many kiloparsecs away, we will usually set our virtual detectors at distances as small as $1000M$ in geometrized units, $M$ being the mass of the source. Also note that $\mathcal{S}$ does not represent a true physical surface, but it is just the transformed integration domain after the change of variables from solid angle to impact parameters. Thus, equation~(\ref{eq:flux_camera_angle}) is not precisely the flux through an extended physical surface, but represents the flux through a surface element at the location of the observer.

In order to compute the integral of equation~(\ref{eq:flux_camera}) as a Riemann sum, we take a grid of $N_\alpha \times N_\beta$ points,
\begin{align}
    \alpha_m &= -\frac{L_\alpha}{2} + \left(m+\frac{1}{2}\right) \Delta \alpha\,,  \quad 0 \leq m \leq N_\alpha-1\,, \\
    \beta_m &= -\frac{L_\beta}{2} + \left(n+\frac{1}{2}\right) \Delta \beta \,,  \quad 0 \leq n \leq N_\beta-1\,,
\end{align}
where $L_\alpha$ and $L_\beta$ are the sides of the image plane, $\Delta \alpha = L_{\alpha}/N_{\alpha}$ and $\Delta \beta = L_{\beta}/N_{\beta}$. The image plane must be large enough to cover the image of the source, usually meaning that it has to be of about the same size as the source (recall that the image plane does not represent a true physical surface).

Then, each grid-point is taken as the initial position of a photon with initial three-momentum normal to the image plane and pointing towards the source. The time component of the four-momenta are fixed so that the resulting four-vector is null, with the choice of the negative sign for the geodesic to be past-directed. When the system we consider is stationarily rotating, it is enough to trace the geodesics for a single common initial time, say $t=0$. Otherwise, we simply have to evolve various photon grids starting at different initial times.

The ray tracing is done as described in Section~\ref{sec:ray_tracing}, with the only difference that in this scheme the equations are solved towards the past. The rays are traced until they intersect the emitting surface or otherwise stray too far away from the source without having hit it.

We know the quantity $\nu^{-3} I_\nu$ is both geodesic and Lorentz invariant, so it can be used to compute $I_\nu(\alpha,\beta,t)$ in terms of the intensity at the source in the local comoving frame where the emission model is defined. The ratio of the frequency at the camera to the frequency at the source in the comoving frame is
\begin{equation}
    \frac{\nu}{\nu_{\text{em}}} = \frac{g_{\mu \nu, \mathrm{i}} k_\mathrm{i}^{\mu} t^{\nu}}{g_{\mu \nu, \mathrm{f}} k_\mathrm{f}^{\mu} u^{\nu}}\,,
\end{equation}
where $g_{\mu \nu,\mathrm{i (f)}}$ is the metric at the initial (final) point of the geodesic, $k_\mathrm{i (f)}^{\mu}$ is the initial (final) four-momentum, $t^\mu = \partial_t$ is the four-velocity of the observer and  $u^{\mu}$ is the local four-velocity of the emitter. For a given photon trajectory, this quotient is independent of the initial frequency. Thus, if a geodesic with initial coordinates on the image plane $(\alpha,\beta)$ intersects the emitting surface at a time $t_{\text{em}}<0$ and spatial position $\mathbf{x}_{\text{em}}$, then
\begin{equation}
    I_\nu(\alpha,\beta,t) = \left(\frac{\nu}{\nu_{\text{em}}}\right)^3 I^{(0)}_{\nu_{\text{em}}}\left(\mathbf{x}_{\text{em}},t+t_{\text{em}}\right)\,,
    \label{eq:image_plane_intensity}
\end{equation}
where $I^{(0)}_{\nu_{\text{em}}}$ is the specific intensity in the comoving frame\footnote{Usually, the astrophysical model provides such an intensity at the source. For example, in spinning neutron stars, solving the local transport in the atmosphere results in a specific intensity distribution over the surface of the star \citep[e.g.][]{heinke2006hydrogen,potekhin2014atmospheres}. However, in other cases the model might instead provide an emissivity with support on a spacetime hypersurface. This special case can still be treated with the techniques described above, but an additional factor dependent on the incidence angle of the ray to the surface must be included. To see why, imagine the hypersurface locally as a thin layer of finite width: clearly, the intensity that a ray picks up when crossing the layer is proportional to the length of the path that it traces within it. Therefore, an additional factor must be $1/n^{\mu}k_{\mu}$, where $n^{\mu}$ is the unit normal to the hypersurface (this is most easily seen in the local comoving frame, where the factor equals $1/\cos\alpha$, and $\alpha$ is the angle between the vectors). The factor remains present after taking the limit as the width tends to zero.}.

Finally, we approximate the flux as
\begin{equation}
    F_\nu(t) \simeq \frac{\Delta \alpha \Delta \beta}{D^2}   \sum_{m,n} I_{\nu}(\alpha_m,\beta_n,t)\,,
    \label{eq:flux_riemann}
\end{equation}
where the sum is over all photons on the grid of the image plane.

\section{Verification of the ray-tracing algorithm}
\label{sec:verification}

\subsection{The Kerr metric in Kerr-Schild Cartesian coordinates}

Even though our code admits arbitrary space-time geometries, our present applications and verification tests circumscribe to the Kerr metric, which we briefly introduce in this section. We have chosen Kerr-Schild Cartesian coordinates, whose benefits are their intuitiveness and their regularity across the black hole event horizon and over the symmetry axis. The downside of these coordinates is that all the components of the metric are nonzero, and the Christoffel symbols are quite involved (see Appendix~\ref{app:appendix} for explicit expressions). Since we envision to use this code for much more complicated space-time geometries, which might not even be given as symbolic functions, we are not particularly worried about these algebraic complexities.

The Kerr spacetime is the unique vacuum stationary black hole solution of Einstein's equations. It is parameterized by two quantities, the mass $M$ and the spin $a$, and it is extremely useful in astrophysical problems. 

In Kerr-Schild Cartesian coordinates (and geometrized units), the Kerr metric takes the form
\begin{equation}
\label{eq:kerrmetric}
g_{\mu\nu}= \eta_{\mu\nu}+ 2 H l_\mu l_\nu\,,
\end{equation}
where $\eta_{\mu\nu}$ is the flat metric, and
\begin{equation}
H=\frac{Mr^3}{r^4+a^2 z^2}\,, \quad
l_\mu=\left(1,\frac{rx+ay}{r^2+a^2},\frac{ry-ax}{r^2+a^2},\frac{z}{r}\right)\,.
\label{eq:lmu}
\end{equation}
The function $r$ is implicitly defined by
\begin{equation}
\frac{x^2+y^2}{r^2+a^2}+\frac{z^2}{r^2}=1\,.
\end{equation}
Any metric in the form of equation~(\ref{eq:kerrmetric}) --for arbitrary $H$ and $l_\mu$-- is said to be in Kerr-Schild form. This metric is stationary and axi-symmetric. For values of $M>0$ and $0 \leq a/M \leq 1$ there is a black hole region in the spacetime. Other values of $a$ are regarded unphysical due to the presence of a naked singularity.
The vector $l^\mu$ is null both with respect to $g_{\mu \nu}$ and $\eta_{\mu \nu}$. Moreover, $l^\mu$ is also geodesic with respect to both metrics, i.e., $l^\mu \partial_\mu l^\nu = l^\mu \nabla_\mu l^\nu = 0$.

In Cartesian coordinates the components of the metric remain regular across the event horizon, which is convenient since it prevents numerical issues to arise close to the black hole region. These coordinates are also regular over the symmetry axis of the metric. Moreover, they also have the appealing property that $\sqrt{-g}=1$ everywhere. Hence, the invariant volume element is homogeneous, thus doing justice to the Cartesian nature of the coordinates. This property is especially useful in the emitter-to-observer scheme, since the invariant volume element for the photon packet sampling is simply $d^4 x$.

\subsection{Conservation of the constants of motion}
\label{sec:constmotion}

\begin{figure}
    \includegraphics[width=\columnwidth]{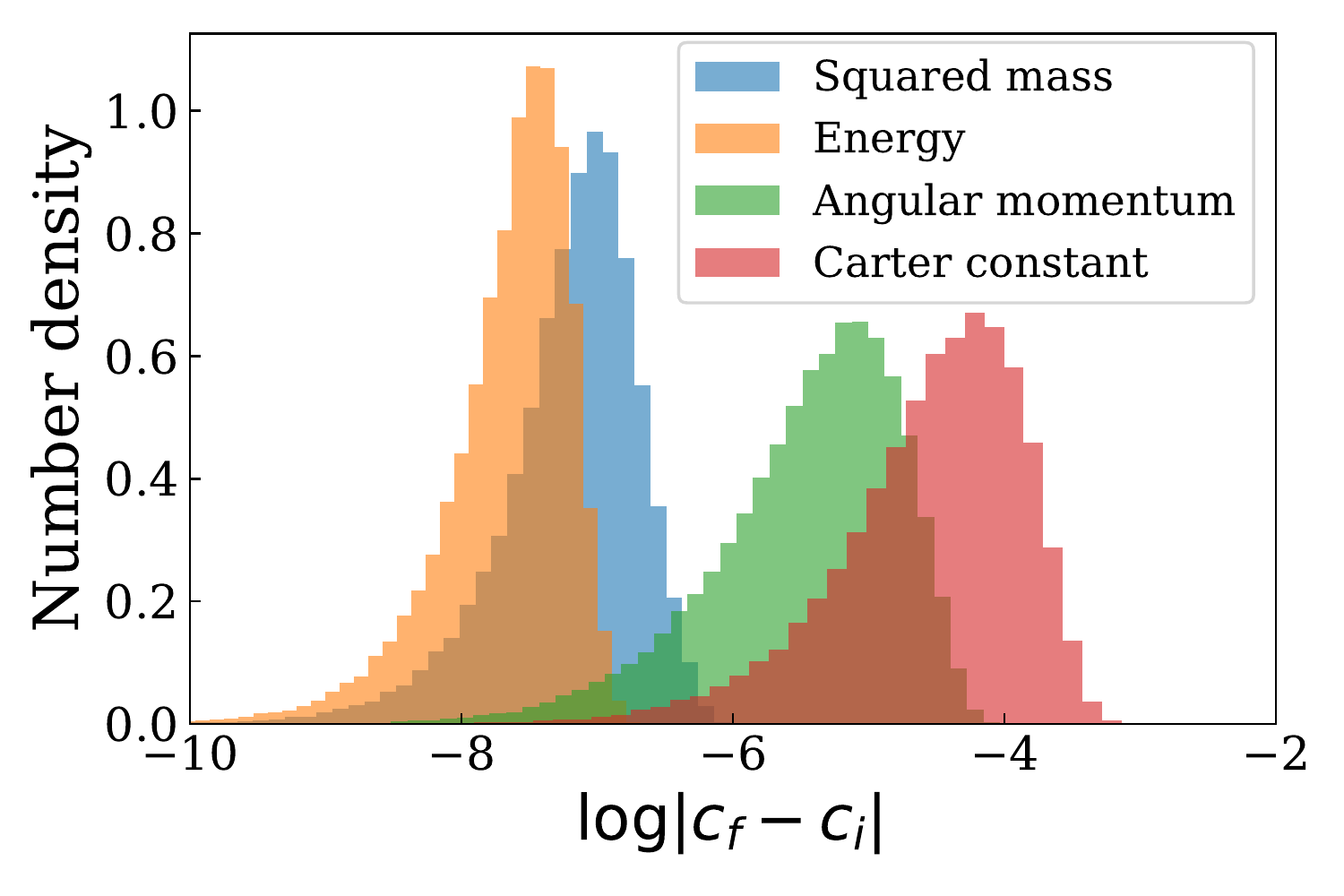}
    \caption{Absolute errors in the conservation of the four constants of motion for initial photons over a meridian at $\rho=5M$ in Kerr spacetime with $a/M=0.99$.}
    \label{fig:absolute-errors}
\end{figure}

\begin{figure}
    \includegraphics[width=\columnwidth]{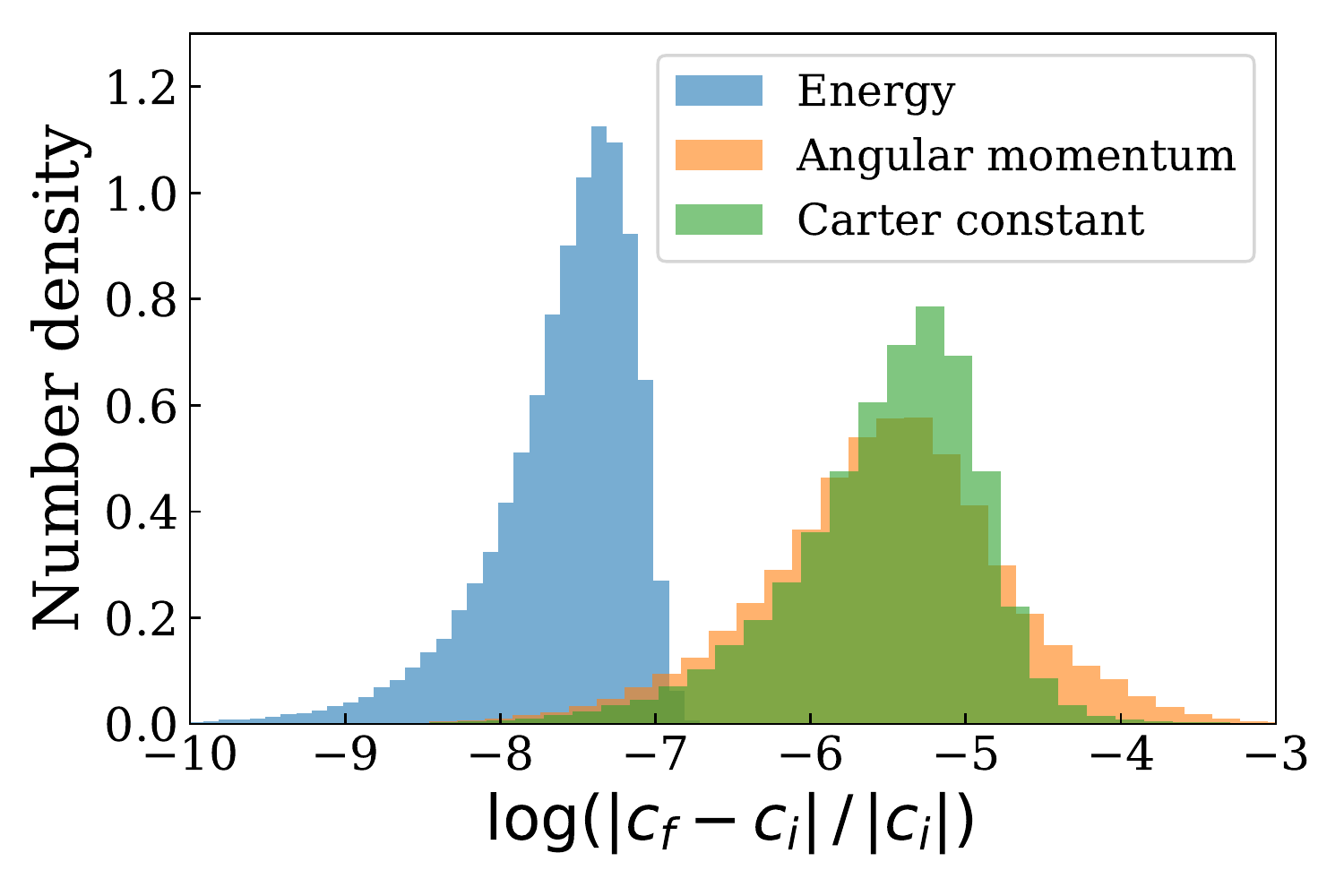}
    \caption{Relative errors in the conservation of the constants of motion for initial photons over a meridian at $\rho=5M$ in Kerr spacetime with $a/M=0.99$. (The exact value of the mass is zero, therefore its relative error is unstable and is disregarded.)}
    \label{fig:relative-errors}
\end{figure}

The Kerr spacetime is stationary and axisymmetric, meaning that it has a time-translation Killing vector $K=\partial_t$, and a rotational Killing vector $R=-y\partial_x + x\partial_y$. These Killing vectors provide two constants of motion: the energy $E=-K_{\mu}k^{\mu}$ and the angular momentum $L=R_{\mu}k^{\mu}$, where $k^{\mu}$ is the tangent vector of the geodesic. The metric itself, as a trivial Killing tensor, provides the constant of motion $m^2=-g_{\mu \nu}k^{\mu}k^{\nu}$, namely the squared mass. The Kerr metric is of Petrov type D, whereas $l^\mu$ is precisely one of its repeated principal null vectors. This results in the presence of a fourth constant of motion, the Carter constant, not related to the obvious symmetries of the metric. The constant comes from an additional Killing tensor, independent of the metric and tensor products of Killing vectors, which can be written as
\begin{align}
C_{\mu\nu}=- s_{(\mu} l_{\nu)} \Delta + r^2 g_{\mu\nu}\,,
\end{align}
where $\Delta=r^2-2Mr+a^2$, $l_\mu$ is the null one-form of equation~(\ref{eq:lmu}), and
\begin{align}
s_\mu = l_\mu + \frac{2a}{\Delta}R_\mu+\frac{2(r^2+a^2)}{\Delta}K_\mu\,.
\end{align}
The Carter constant then reads
\begin{multline}
C = C_{\mu \nu}k^{\mu}k^{\nu} = -\Delta (l_\mu k^\mu)^2 \\ -2l_\mu k^\mu [aL-(r^2+a^2)E]-r^2 m^2\,.
\end{multline}
We used these four constants of motion as a test of the accuracy of our ray-tracing algorithm. As an example we set $100$ initial points over a meridian of the sphere $\rho^2 = x^2+y^2+z^2=(5M)^2$ in Kerr spacetime with $a/M=0.99$. We took $10^3$ photons at each point, with directions of emission isotropically distributed over the outward directed hemisphere, yielding a total $N=10^5$ photons. We propagated these photons up to $r=10^3M$ using the method "VCABM" (see Section~\ref{sec:ray_tracing}). The relative tolerance of the method was set to $10^{-8}$. In Fig.~\ref{fig:absolute-errors} and Fig.~\ref{fig:relative-errors} we show the results obtained for the absolute and relative errors in the conservation of the constants of motion between the initial and final points. (We exclude the relative error in $m^2$ because its exact value is close to zero). Naturally, the particles, which are initially massless, acquire mass due to numerical errors in the evolution. However, these differences are very small, and the conformity with the expected accuracy is excellent.

\subsection{Comparison with the ray-tracing function}

Some codes which are restricted to the Schwarzschild spacetime have used what is called the ray-tracing function to compute the deflection angle of geodesics in a semianalytic manner \citep[e.g.][]{pechenick1983hot, page1994surface, perna2008constraints}. That expression can be used as a verification of our ray-tracing alogrithm. 

Let a photon be located at radius $r$, and let $\delta$ be the angle between the three-momentum of the photon and the radial vector $\partial_r$ measured in the frame of a static observer. The angle $\delta$ satisfies
\begin{equation}
\sin\delta = \frac{b}{r} \sqrt{1-\frac{2M}{r}}\,,
\end{equation}
where $b=L/E$ is the impact parameter of the photon. Then, the deflection angle can be written as  
\begin{multline}
\label{eq:ray-tracing-integral}
\alpha_{\delta}=\int_0^{\frac{R_s}{2r}} \mathop{du}  \sin\delta \left[ \left(1-\frac{R_s}{r}\right) \left(\frac{R_s}{2r}\right)^2 \right. \\ \left. -\left(1-2u \right) u^2 \sin^2\delta  \right]^{-1/2}  -\delta\,,
\end{multline}
where $R_s=2M$ is the Schwarzschild radius. 

\begin{figure}
    \includegraphics[width=\columnwidth]{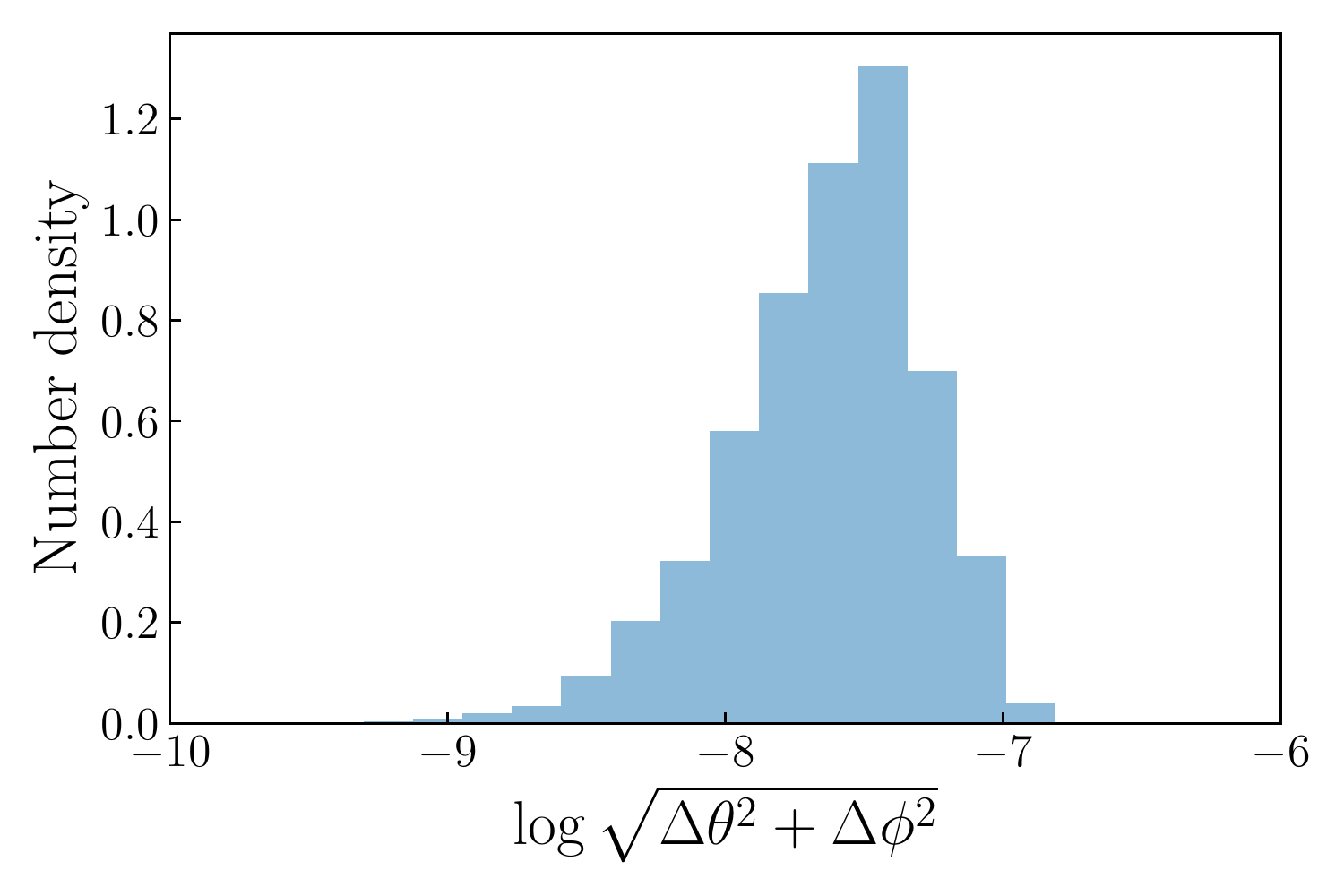}
    \caption{Difference between the final angles obtained with our code and those obtained with the ray-tracing function of equation~(\ref{eq:ray-tracing-integral}) for a total of $10^4$ photons initially at $(x,y,z)=(0,0,5M)$ in Schwarzschild spacetime and with directions isotropically distributed on the outward pointing hemisphere.}
    \label{fig:deflection-angle}
\end{figure}

As a test of our ray-tracing algorithm, we took $10^4$ photons at $(x,y,z)=(0,0,5M)$ with directions of emission isotropically distributed over the outwards directed hemisphere. We propagated those photons up to $r=10^3M$ using the method "VCABM" (Section~\ref{sec:ray_tracing}). The relative tolerance of the method was set to $10^{-8}$. In Fig.~\ref{fig:deflection-angle} we show the difference $\sqrt{\Delta \theta^2 + \Delta \phi^2}$ between the final angles obtained with our code and the final angles computed from formula~(\ref{eq:ray-tracing-integral}). Again, the agreement is excellent. 

\subsection{Interpolation of a numerical metric}

In principle, \texttt{Skylight} can also handle numerical --i.e. tabulated-- metrics. Although our applications so far restrict to analytical metrics, to test this possibility in a simple setting we implemented the metric of a general spherically symmetric spacetime, which in spherical coordinates takes the form
\begin{equation}
    ds^2 = -e^{2\alpha(r)} dt^2 + e^{2\beta(r)} dr^2 + r^2 (d\theta^2 + \sin^2 \theta d\varphi^2)\,,
\end{equation}
where $\alpha(r)$ and $\beta(r)$ are arbitrary functions that might be provided in the form of tables. The Christoffel symbols can be computed straightforwardly from the expression above, and the code does this before starting the integration, to avoid having to compute many numerical derivatives at each time step. Once the metric and the Christoffel symbols are tabulated, we build interpolator functions for them. 

\begin{figure}
    \includegraphics[width=\columnwidth]{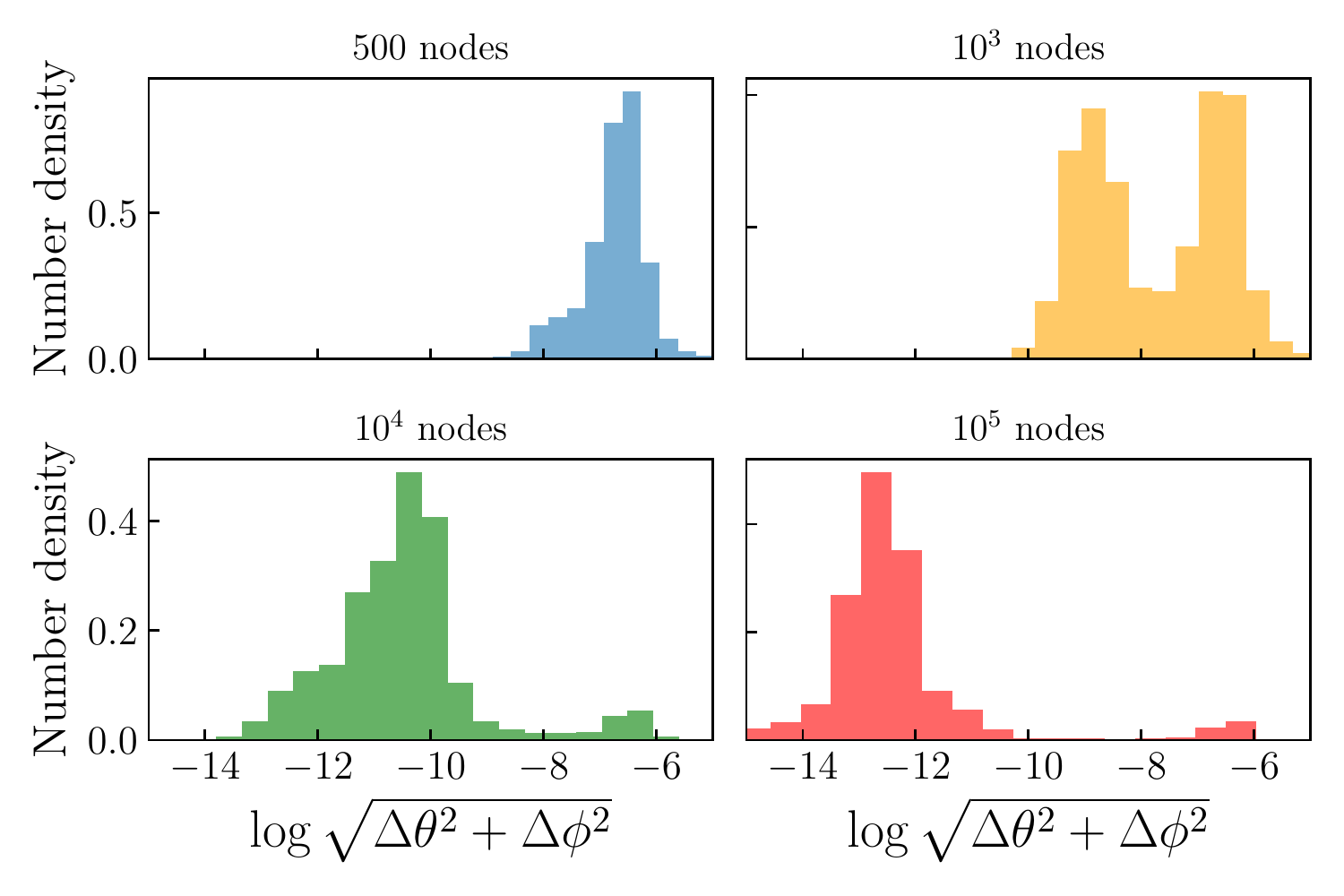}
    \caption{Difference in final spherical angles for the evolution of $10^4$ photons initially at $r=5M$, with the interpolated Schwarzschild metric and the analytical metric. Each panel corresponds to different resolution of the numerical metric, with $500$, $10^{3}$, $10^{4}$ and $10^{5}$ logarithmically spaced nodes correspondingly.}
    \label{fig:deflection-angle_interp}
\end{figure}

These spacetimes correspond to spherically symmetric distributions of matter. In the case of vacuum in GR, the metric reduces to the Schwarzschild metric, with
\begin{align}
    \alpha(r) &= \frac{1}{2} \log \left(1-\frac{2M}{r}\right)\,, \\
    \beta(r) &= -\alpha(r)\,.
\end{align}
Therefore, we can use mock tables of the Schwarzschild metric and connection functions to test our interpolation scheme for ray tracing in numerical spacetimes. For this test, we applied the same procedure as in the previous section: as initial data, we took $10^{4}$ photons at $r=5M$ with emission directions isotropically distributed outwards and we propagated those photons up to $r=10^3M$ both with the analytical and numerical metrics. We used four distinct resolutions for the metric tables, with $500$, $10^{3}$, $10^{4}$ and $10^{5}$ logarithmically spaced nodes between $r=2.1M$ (close to the event horizon) and $r=10^{3} M$. The relative tolerance of the geodesic integration method was set to $10^{-8}$. In Fig.~\ref{fig:deflection-angle_interp} we show the difference $\sqrt{\Delta \theta^2 + \Delta \phi^2}$ for the final spherical angles in all four cases with respect to the angles obtained with the analytical metric. In all cases the matching is excellent, and it improves with increasing resolution.

\section{Astrophysical tests}
\label{sec:astrophysical_tests}

\subsection{Relativistically broadened emission line from a thin accretion disk}

The gravitational influence of accreting black holes has important effects on the emission generated on their accretion disks. In particular, the iron emission lines produced in such accretion disks might suffer a relativistic broadening due to the combined effect of gravitational redshift and Doppler boosting to the frame of the fluid. Here we reproduce the results of \citet{dexter2009fast} for the broadening of an emission line in a simple model of a thin accretion disk around a Kerr black hole. The disk is optically thick and geometrically thin and lies on the equatorial plane of the black hole. The inner radius of the disk is defined as the radius of the marginally stable circular orbit $r_{\mathrm{ms}}$ \citep{bardeen1973timelike}, and the outer radius is $r_{\text{out}} = 15M$. The particles of the disk rotate in circular orbits, with an angular speed given by 
\begin{equation}
    \omega_{\pm} = \frac{\pm \sqrt{M}}{r^{3/2} \pm a \sqrt{M}}\,,
\end{equation}
where $\omega_+$ and $\omega_{-}$ correspond to prograde and retrograde disks respectively. The emissivity is defined in the local comoving frame of the disk, and it is monochromatic, isotropic, and weighted by a factor of $r^{-2}$:
\begin{equation}
    j_{\nu}(x^{\mu}) \propto \frac{1}{r^2} \delta(z) \delta(\nu-\nu_0) \chi(r) \,,
\end{equation}
where the function $\chi(r)$ equals $1$ if $r_{\mathrm{ms}}\leq r \leq r_{\mathrm{out}}$ and zero otherwise, and $\nu_0$ is the frequency of emission in the comoving frame.

\begin{figure}
    \includegraphics[width=\columnwidth]{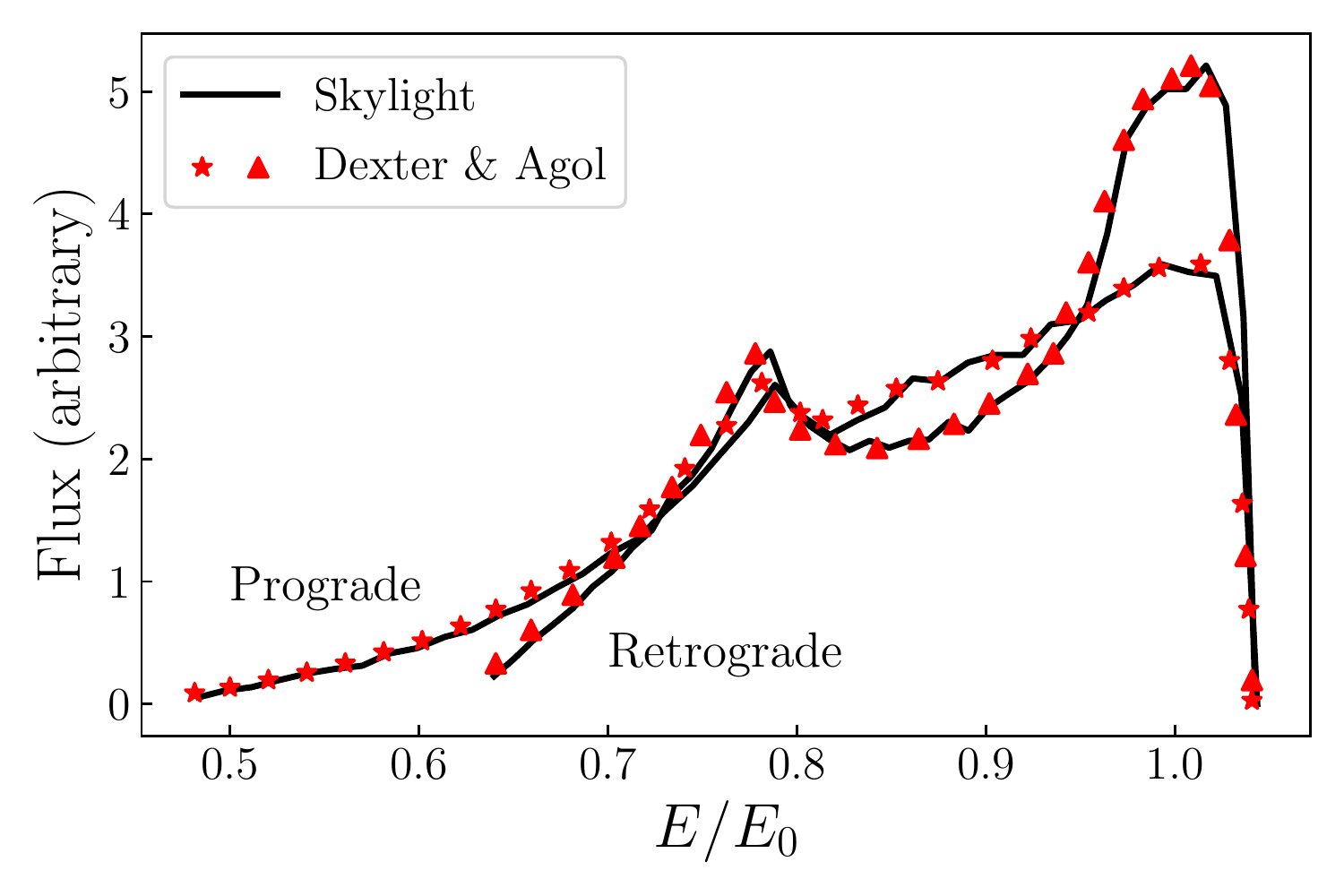}
    \caption{Relativistic broadening of an emission line in a thin accretion disk with the observer-to-emitter scheme of \texttt{Skylight}. The black hole spin is $a/M=0.5$ and the viewing angle is $\xi=30^{\circ}.$}
    \label{fig:line-broadening-reverse}
\end{figure}

\begin{figure}
    \includegraphics[width=\columnwidth]{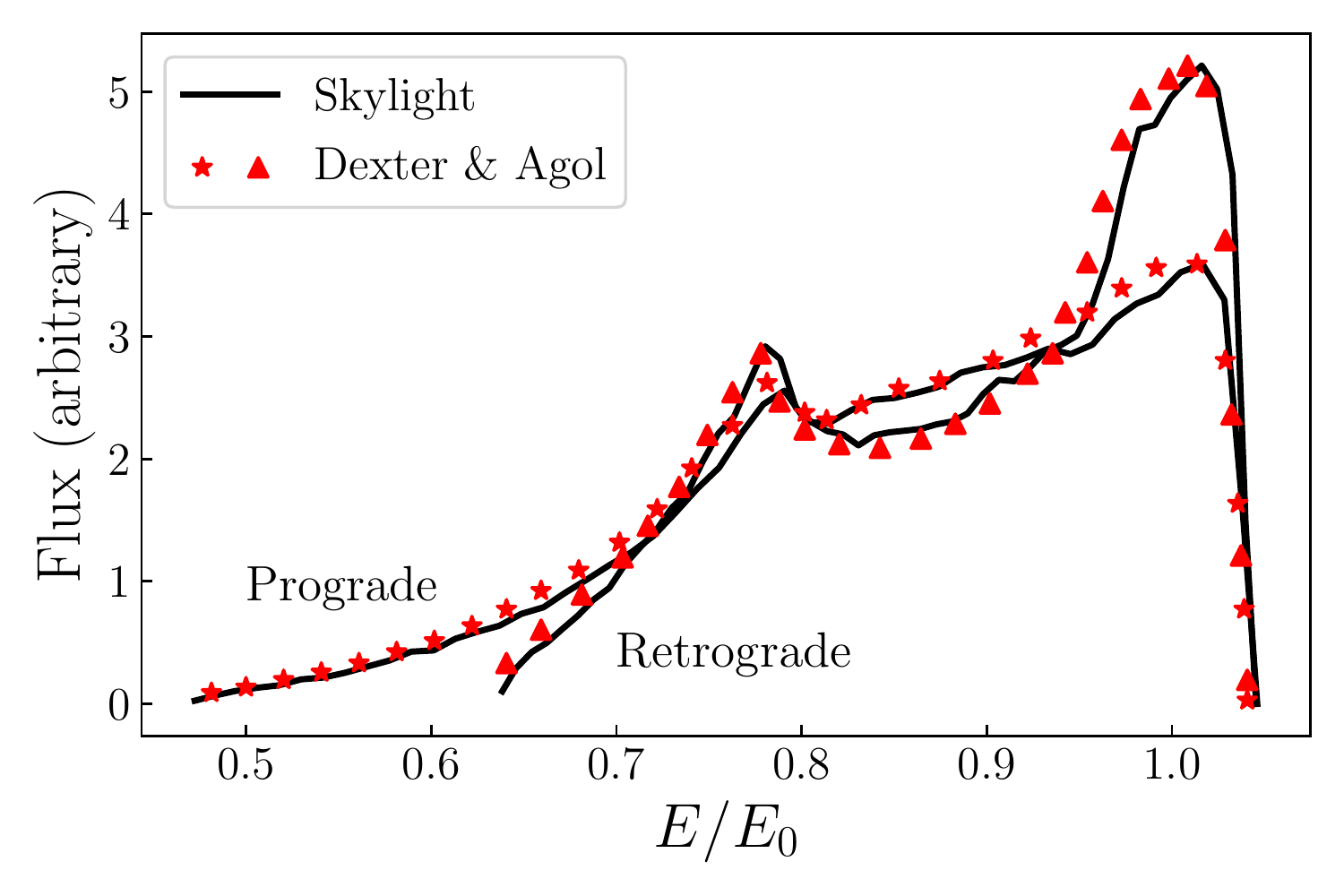}
    \caption{Relativistic broadening of an emission line in a thin accretion disk with the emitter-to-observer scheme of \texttt{Skylight}. The black hole spin is $a/M=0.5$ and the viewing angle is $\xi=30^{\circ}$.}
    \label{fig:line-broadening-direct}
\end{figure}

\begin{figure}
    \includegraphics[width=\columnwidth]{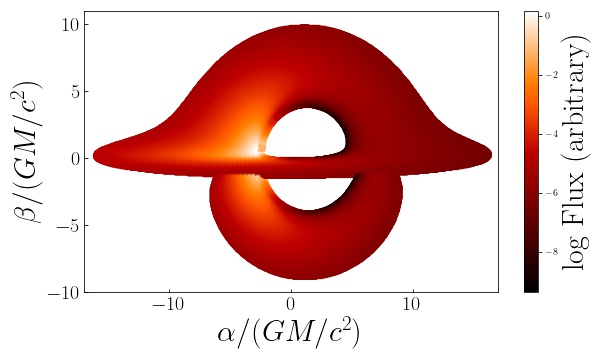}
    \caption{Image of the simple thin disk model with black hole spin $a/M=0.99$ with the observer-to-emitter scheme at a viewing angle of $\theta=85^{\circ}$.}
    \label{fig:bh_image_reverse}
\end{figure}

We computed the spectrum of the disk for an observer at a distance $r=10^3M$ and an inclination angle of $\xi=30^{\circ}$ in the cases of prograde and retrograde rotation using both schemes of \texttt{Skylight}. The black hole spin is $a/M=0.5$. In the observer-to-emitter scheme, we used a square image plane of side $L=2.1r_{\mathrm{out}}$ at a distance $d=10^3M$ with $N=200$ grid points per side. In the emitter-to-observer scheme, we took $\num{5e-3}$ initial points on the disk and $\num{5e-3}$ uniformly sampled emission directions in the comoving frame, with a total of $N=\num{2.5e7}$ photons. The virtual detectors are located at $d=10^3M$. In Fig.~\ref{fig:line-broadening-reverse} and Fig.~\ref{fig:line-broadening-direct} we compare the results of both schemes to those of \cite{dexter2009fast}, finding an excellent agreement in both cases. Also, in Fig.~\ref{fig:bh_image_reverse} we show an image of the disk model for a black hole spin of $a/M=0.99$ as seen at an inclination angle of $\xi = 85^{\circ}$ obtained with observer-to-emitter scheme of \texttt{Skylight} using the same image plane size as above but increasing the resolution to $N=500$ grid points per side of the image plane.

\subsection{Hot spot orbiting a Schwarzschild black hole}
\label{sec:hot_spot}
Since the accretion disk of the previous test is stationary and axisymmetric, it serves mostly as a probe of the spectral dependence of \texttt{Skylight} but not as much of timing. To test the correct treatment of timing in our code we implemented a different model: the orbiting hot spot model described in \cite{schnittman2004harmonic}. The emission region is a circular spot orbiting a Schwarzschild black hole on the equatorial plane. The center of the spot follows the innermost stable retrograde circular orbit with radius $r = 6M$ at a retrograde Keplerian angular speed $\omega = -\sqrt{M}/r^{3/2}$. The emissivity is monochromatic and isotropic in the frame of the spot, and is modulated by a Gaussian profile
\begin{equation}
    j_{\nu}(\mathbf{x^{\mu}}) \propto \delta(z) \delta(\nu-\nu_0) \exp \{-|\mathbf{x}-\mathbf{x}_{\mathrm{spot}}(t)|^2 / 2 R_{\mathrm{spot}}^2 \} \,,
\end{equation}
where $\mathbf{x}=(x,y,z)$, $\mathbf{x}_{\mathrm{spot}}(t)$ is the position of the spot center, $\nu_0$ is the frequency of emission in the local comoving frame of the spot, and the standard deviation of the Gaussian profile is $R_{\mathrm{spot}}=0.25M$. Due to the small size of the spot, the distance of an emission point to the center of the spot is computed as in local Euclidean geometry. In practice, we truncate the emissivity at a distance $4 R_{\mathrm{spot}}$ from the center of the spot, where it is safely close to zero. The four-velocity of all points inside the spot is taken equal to that of the guiding geodesic trajectory. This means that the energy in the local "comoving" frame of the spot of a photon with four-momentum $k^{\mu}$ located anywhere inside the spot is calculated as $-k_\mu v^{\mu}(\mathbf{x}_\mathrm{spot})$, where $v^{\mu}(\mathbf{x}_\mathrm{spot})$ is the four-velocity of the guiding geodesic trajectory. 

\begin{figure}
    \includegraphics[width=\columnwidth]{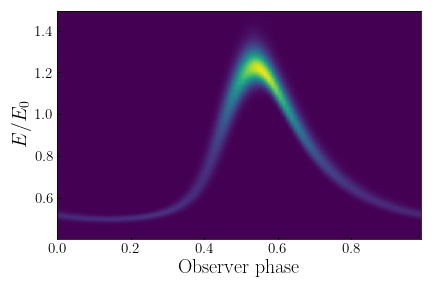}
    \caption{Spectrogram of a circular hot spot of radius $R=0.25M$ for a viewing angle of $i=60^{\circ}$ in the observer-to-emitter scheme.}
    \label{fig:hot_spot_spec}
\end{figure}

\begin{figure}
    \includegraphics[width=\columnwidth]{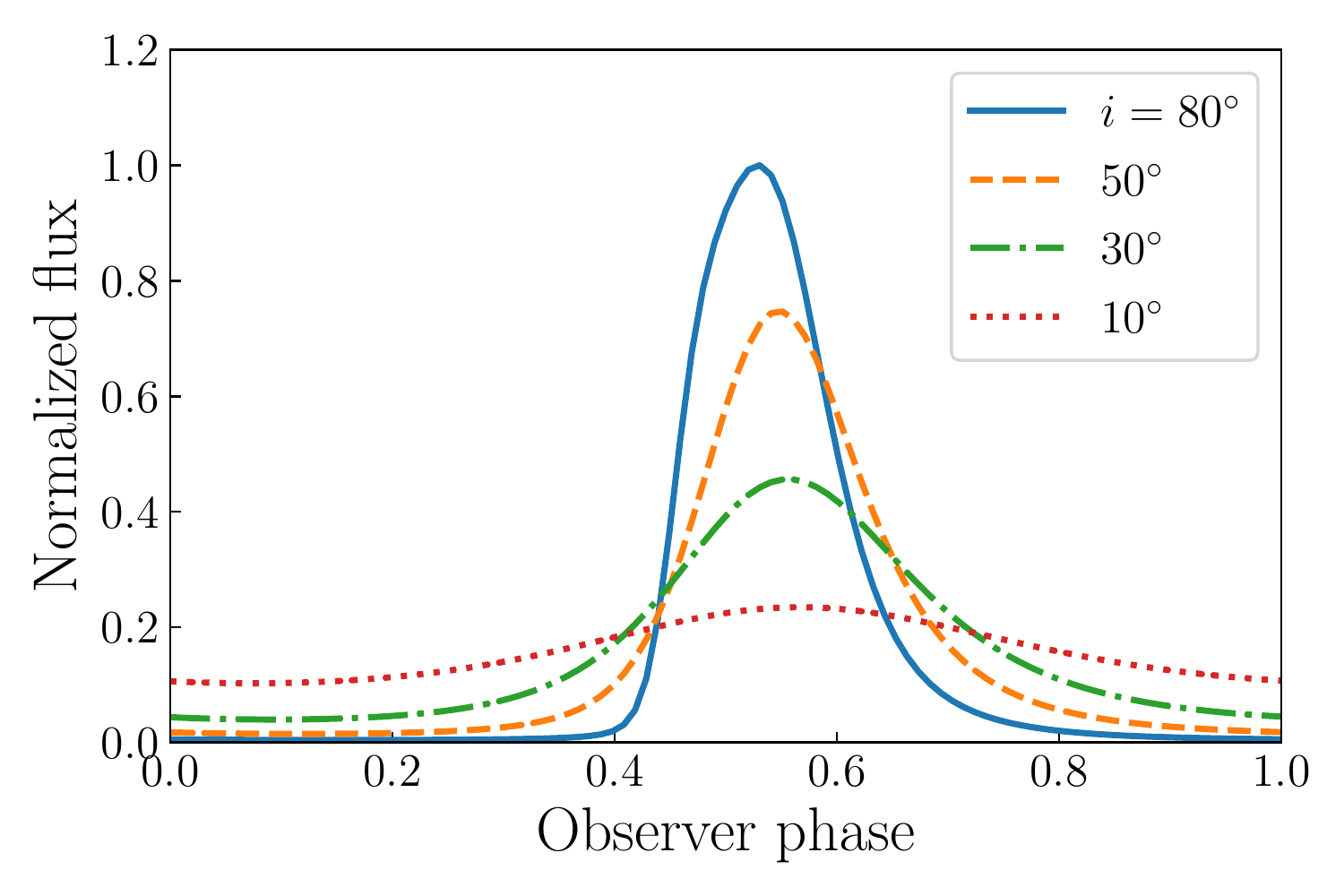}
    \caption{Light curves for a circular hot spot of radius $R=0.25M$ at various viewing angles using the observer-to-emitter scheme. The light curves are normalized to $1$ and scaled to the maximum value of the curve at $i=80^{\circ}$.}
    \label{fig:hot_spot_cam}
\end{figure}

For the observer-to-emitter scheme we used square image planes of side $L=20M$ at a distance $d=10^3M$ with $N=800$ grid points per side. In Fig.~\ref{fig:hot_spot_spec} we show a spectrogram obtained with this scheme for a viewing angle of $i=60^{\circ}$ and in Fig.~\ref{fig:hot_spot_cam} we show the bolometric light curves at various viewing angles. Each light curve is normalized to $1$ and then scaled to the maximum value of the $i=80^{\circ}$ light curve. As the inclination increases, the light curves become more sharply peaked because Doppler beaming becomes more important and gravitational redshift becomes stronger for rays coming from behind the black hole. Both figures are in excellent agreement with the results of \cite{schnittman2004harmonic}.

For the emitter-to-observer scheme, in order to avoid setting many initial photon sets at different locations of the spot along its orbit, we introduce a slight modification to the model. This is because a constant four-velocity throughout the spot is in conflict with the hypothesis of stationarity, i.e. that all physical quantities depend on $\phi$ and $t$ only via the combination $\omega t - \phi$. To simplify the calculations, we must have a self-consistent stationary model which respects that symmetry. Notice that for the spot to maintain its shape along its orbit, the four-velocity cannot be constant over the spot. Thus, we implemented a modified model in which the four-velocities inside the spot correspond to a rigidly rotating hot spot, with the same angular velocity as that of the guiding geodesic trajectory of the original model. These are the four-velocities we use to set the local orthonormal frame at each point. Also, we reduce the radius of the spot to $R_{\mathrm{spot}}=0.05M$ to highlight the region where both four-velocity models approximately agree. 

\begin{figure}
    \includegraphics[width=\columnwidth]{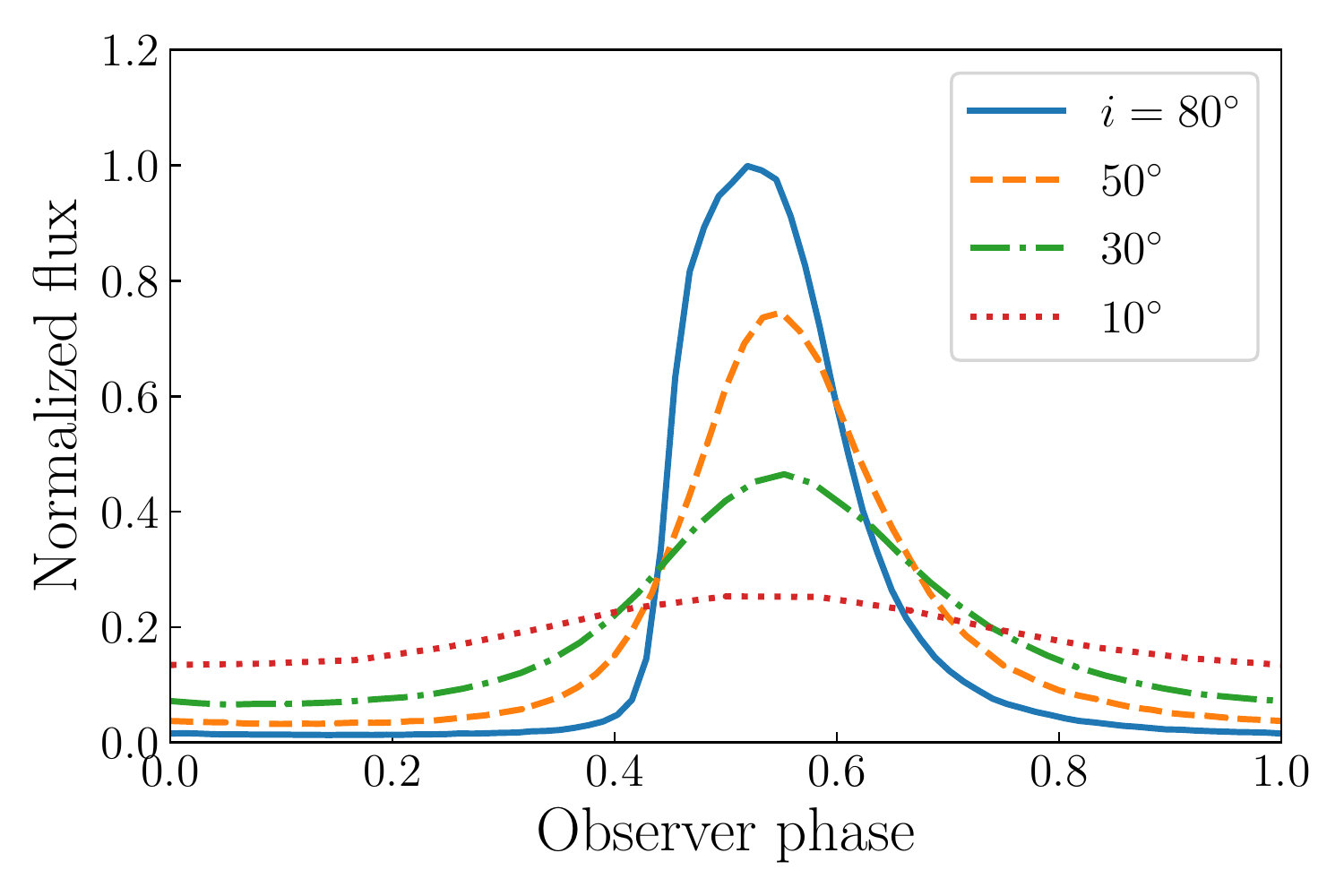}
    \caption{Light curves for a circular hot spot of radius $R=0.05M$ at various viewing angles using the emitter-to-observer scheme. The light curves are normalized to $1$ and scaled to the maximum value of the curve at $i=80^{\circ}$.}
    \label{fig:hot_spot_direct}
\end{figure}

To check consistency, we computed the light curves of this modified model using the observer-to-emitter scheme, finding almost no difference with Figs.~\ref{fig:hot_spot_spec} and \ref{fig:hot_spot_cam} of the original model. Thus, we conclude the modification to the model is not important, and it is still useful to do a comparison of the emitter-to-observer scheme with the results of \cite{schnittman2004harmonic}. This is because, in the end, what only matters is what happens really close to the center of the spot, where both models are essentially the same. Finally, we ran the emitter-to-observer scheme in the modified model with a total $N=128$ million photons propagated from the surface of the spot to the virtual detectors located at $d=10^3M$. In Fig.~\ref{fig:hot_spot_direct} we show the light curves of this modified model for various viewing angles. The light curves are scaled in the same manner as in the observer-to-emitter scheme. Again, the agreement both with the results of the other scheme and with those of \cite{schnittman2004harmonic} is very good. Although the light curve at $i=80^{\circ}$ is somewhat wider than in the other scheme, we consider the similitude to be acceptable, rather focusing on the fact that the relations of the amplitudes of the pulses and the phases of the peaks are correct. Other noticeable differences appear for the lowest inclination light curves. This is due to the fact that the photon collector area gets smaller by a factor of $\sin i$. Therefore, the photon statistics at lower inclinations is expected to be worse as compared to higher inclinations for the same data.

In general, if we are interested in the light curves at only a few inclination angles, using the observer-to-emitter scheme for each angle is still more efficient than using the emitter-to-observer scheme to extract them all at once. However, our main motivation in introducing and testing the emitter-to-observer scheme is that it will be more easily adaptable for the inclusion of scattering processes in the future as a Monte Carlo simulation, rather than trying to include them as emissivity and absorptivity coefficients into the transport equation. Moreover, it is more natural to use the emitter-to-observer scheme for defining astrophysical models in which the location of the emission region is dynamical, as in, e.g., the radio emission from dynamical current sheets obtained with FF simulations of black hole and neutron star magnetospheres.

\subsection{Neutron star hot spot X-ray emission}
\label{sec:bogdanov}

\begin{figure}
    \includegraphics[width=\columnwidth]{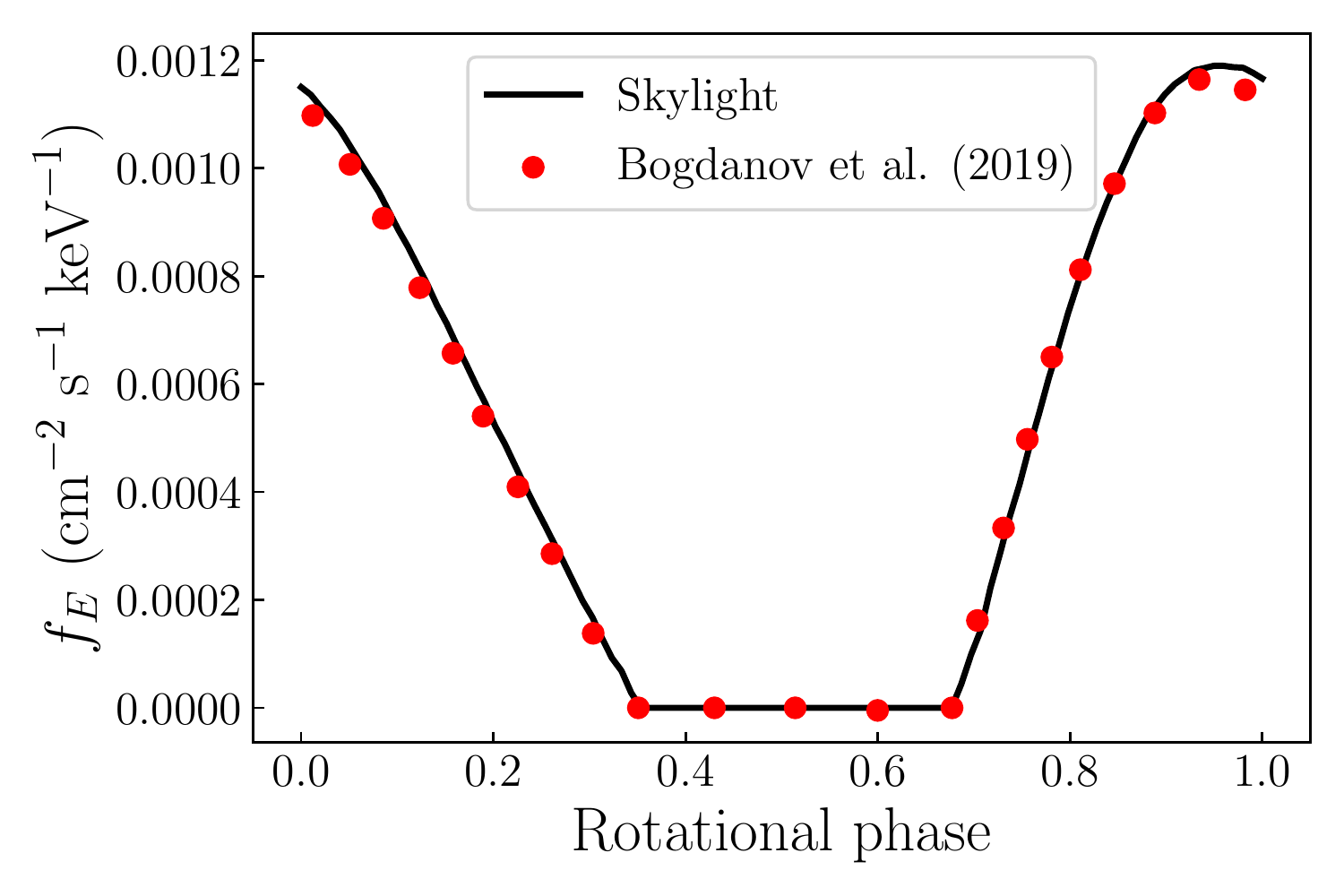}
    \caption{Monochromatic particle flux at $\SI{1}{\keV}$ for the neutron star hot spot model SD1c of Bogdanov et al. (2019). The parameters are $\theta_c=90^{\circ}$, $\Delta \theta=0.01$, $\xi=90^{\circ}$ and $\nu=\SI{200}{\Hz}$.}
    \label{fig:bogdanov_c}
\end{figure}

\begin{figure}
    \includegraphics[width=\columnwidth]{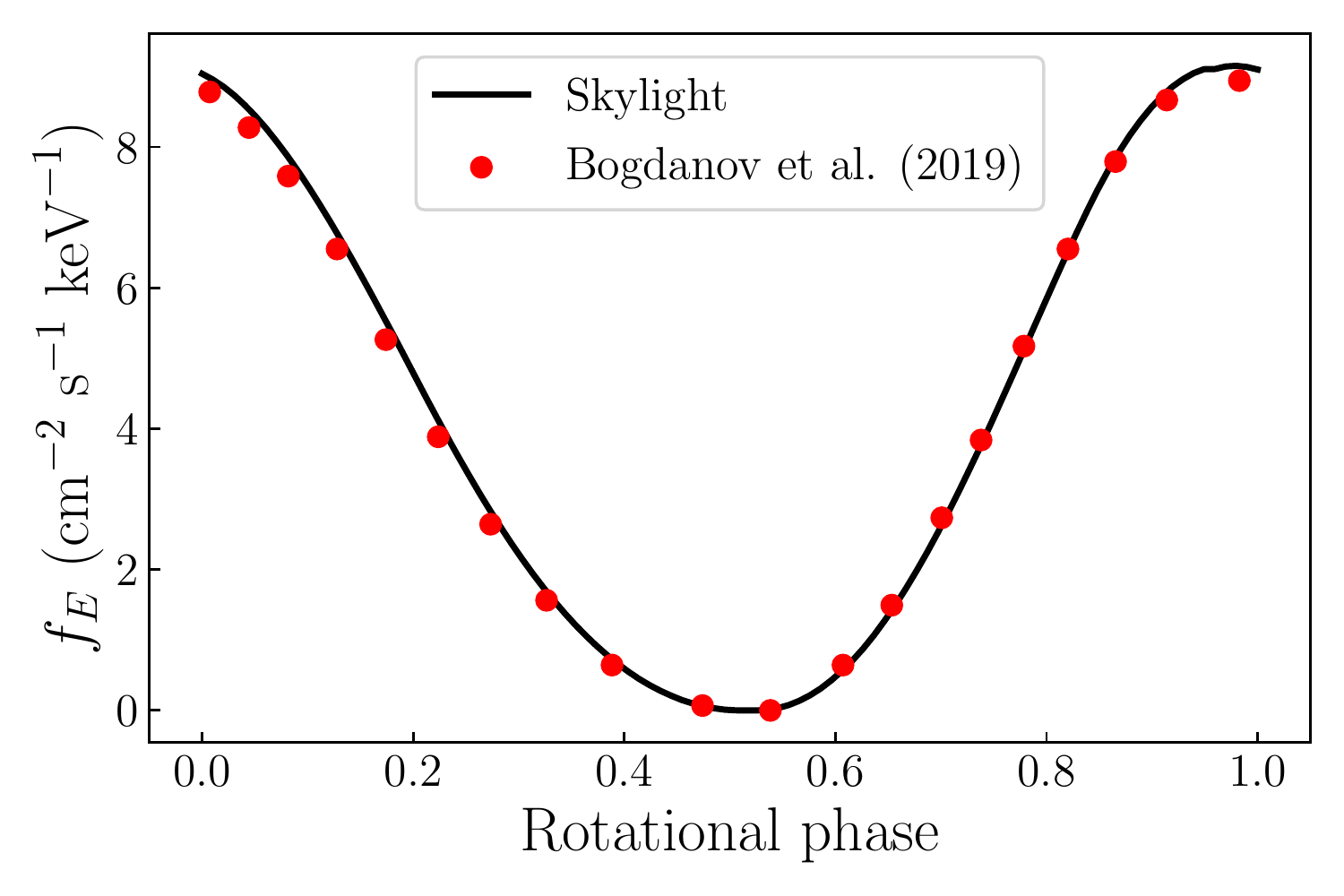}
    \caption{Monochromatic particle flux at $\SI{1}{\keV}$ for the neutron star hot spot model SD1d of Bogdanov et al. (2019). The parameters are $\theta_c=90^{\circ}$, $\Delta \theta=1$, $\xi=90^{\circ}$ and $\nu=\SI{200}{\Hz}$.}
    \label{fig:bogdanov_d}
\end{figure}

\begin{figure}
    \includegraphics[width=\columnwidth]{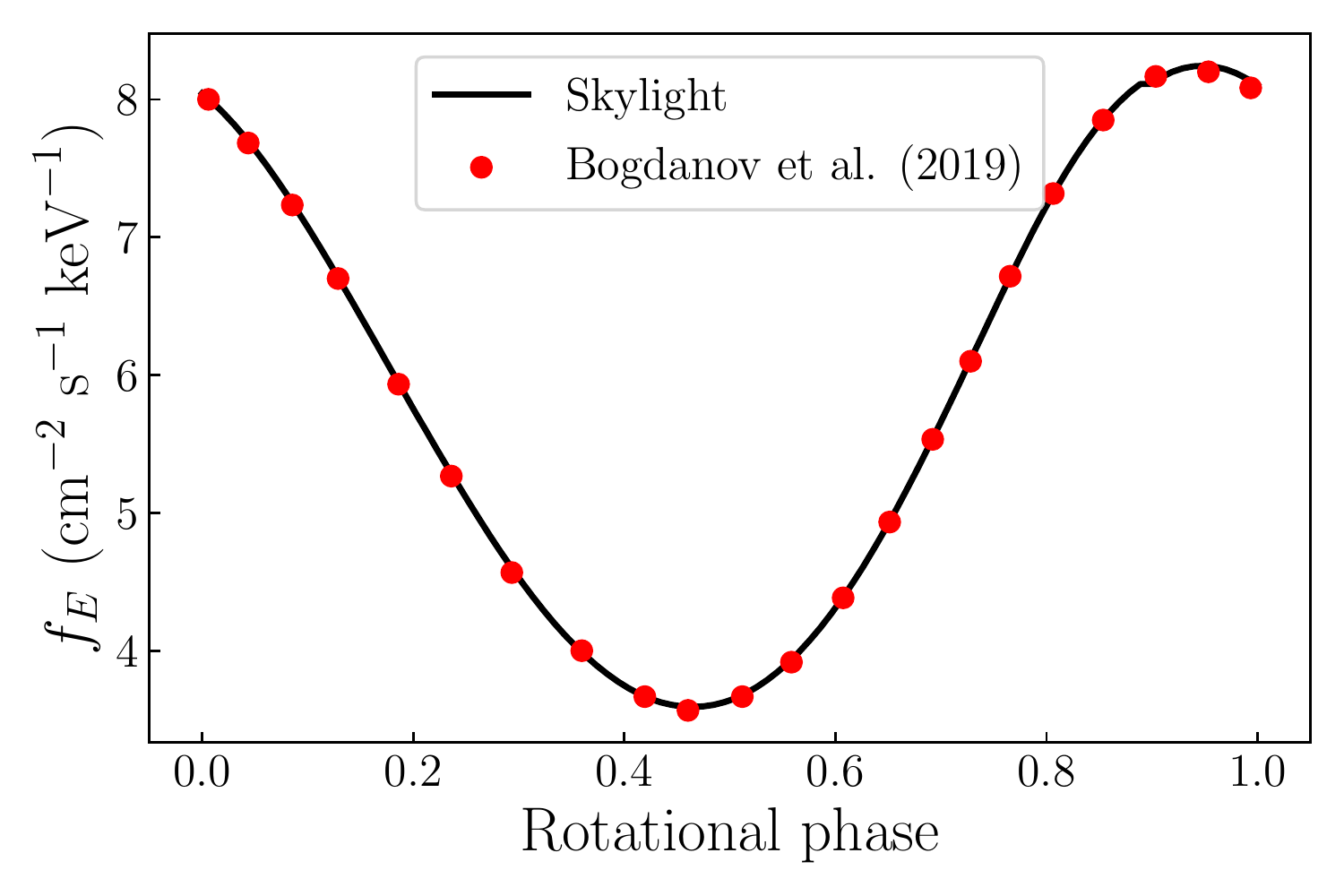}
    \caption{Monochromatic particle flux at $\SI{1}{\keV}$ for the neutron star hot spot model SD1e of Bogdanov et al. (2019). The parameters are $\theta_c=60^{\circ}$, $\Delta \theta=1$, $\xi=60^{\circ}$ and $\nu=\SI{400}{\Hz}$.}
    \label{fig:bogdanov_e}
\end{figure}

\cite{bogdanov2019constraining} provided a set of verified high-precision synthetic neutron star X-ray pulse profiles for other codes to be tested against. We implemented some of those tests, implying a simultaneous test of timing and spectral dependence of \texttt{Skylight}. Besides, the reference pulse profiles are in physical units, therefore they also serve as a test of the normalization of our curves and our treatment of units in general. The multiple codes compared in \cite{bogdanov2019constraining} use the Schwarzschild + Doppler (S+D) and Oblate-Schwarzschild (OS) approximations, in which spacetime is modeled as Schwarzschild and the neutron star surface is supposed to be either a sphere or an oblate spheroid, respectively (see their paper for details). Following their nomenclature, we have implemented the class SD1 of models in which the neutron star is approximated as a spinning sphere in Schwarzschild spacetime. The neutron star's mass and radius are $M=1.4 M_{\odot}$ and $R=\SI{12}{\km}$ respectively. The emission comes from a single circular hot spot on the surface of the star. The specific intensity is assumed to be isotropically distributed and follows a Planckian distribution with $kT = \SI{0.35}{\keV}$ (everything referred to the local corotating frame). The distance to the observer is $D= \SI{200}{\parsec}$. The rest of the parameters of the model are the colatitude of the spot center $\theta_c$, the angular radius of the spot $\Delta \theta$, the colatitude of the observer $\xi$, and the rotation frequency of the neutron star $\nu$. The values of these parameters for the cases we reproduced (SD1c--e) are listed in Table~\ref{tab:params}. In Figs.~\ref{fig:bogdanov_c}--\ref{fig:bogdanov_e} we compare the observed monochromatic particle flux at $\SI{1}{\keV}$ obtained with the observer-to-emitter scheme of \texttt{Skylight} against the profiles of \cite{bogdanov2019constraining}, finding a very good agreement between the profiles.

 \begin{table*}[t]
 \centering
   \caption{Parameters of the reference neutron star hot spot tests we implemented}
  \label{tab:params}
    \begin{tabular}{||c c c c||} 
 \hline
 Parameter & Test SD1c & Test SD1d & Test SD1e \\ [0.5ex] 
 \hline\hline
 Colatitude of the spot center ($^\circ$) & $90$ & $90$ & $60$ \\ 
 \hline
 Angular radius of the spot (rad) & $0.01$ & $1$ & $1$ \\
 \hline
 Colatitude of the observer ($^\circ$) & $90$ & $90$ & $30$ \\
 \hline
 Rotation frequency ($\si{\Hz}$) & $200$ & $200$ & $400$ \\
 \hline
 
\end{tabular}
\end{table*}

\section{Convergence tests}
\label{sec:convergence}

\subsection{Observer-to-emitter scheme}

\begin{figure}
    \includegraphics[width=\columnwidth]{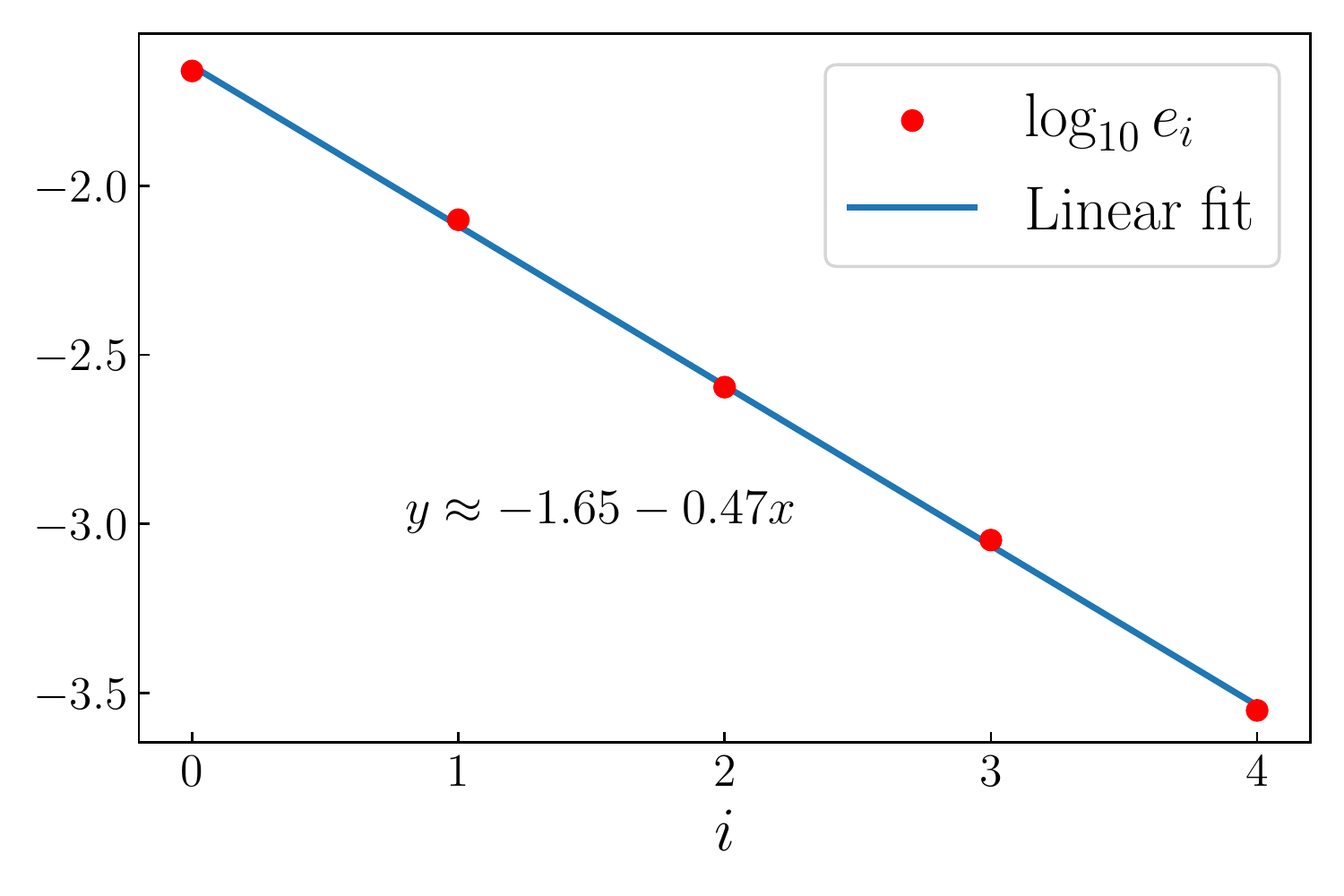}
    \caption{The dots correspond to the relative $L_2$-errors for the convergence test using the SD1d model of Bogdanov et al. (2019) with $N_i = 25 \times 2^i$ points per side of the image plane. The solid line is the least squares linear fit of the data points. The error approximately follows a power law with index $p = -1.57$ in terms of the number of points per side of the image plane.}
    \label{fig:convergence}
\end{figure}

We performed a convergence test for the observer-to-emitter scheme in the simple hot spot model SD1d of \cite{bogdanov2019constraining} described in the Section~\ref{sec:bogdanov}. The observer is at a distance $D= \SI{200}{\parsec}$ and a colatitude $\xi = 90^{\circ}$. The neutron star has a mass of $M=1.4 M_{\odot}$, a radius of $R=\SI{12}{\km}$, and a rotation frequency of $\nu= \SI{200}{\Hz}$. The colatitude of the spot center on the star is $\theta_c = 90^{\circ}$ and its angular radius is $\Delta \theta = 1$. In the corotating frame the emission follows Planck's law with a temperature corresponding to $kT = \SI{0.35}{\keV}$. We compared six different runs using a square image plane of side $L \approx 2.75 R$ and $N_i = 25 \times 2^i$ points per side ($0 \leq i \leq 5$). In Fig.~\ref{fig:convergence} we show the relative $L_2$-errors for the monochromatic particle flux at $\SI{1}{\keV}$ with respect to the highest-resolution run, namely
\begin{equation}
    e_{i} = \frac{ || f_i(t) - f_5(t)||_{L_2}}{||f_5(t)||_{L_2}}\,,\quad 0 \leq i \leq 4\,.
\end{equation}
A least-squares linear fit of the error data gives the relation $\log_{10}(e_i) \approx -1.65 - 0.47 i$, meaning the error approximately follows a power law 
\begin{equation}
e_i \approx \num{8e-4} \left( \frac{N_i}{200}\right)^{-1.57}\,
\end{equation}
in terms of the number of points per side of the image plane.

In other words, for this configuration we find that a resolution of $N = 200$ points per side is enough to obtain an approximate relative error of order $10^{-4}$, and that this error scales as a power law of index $p \approx -1.57$ with respect to $N$. The resolution required for achieving the same error will be greater if the emission region has a higher a complexity or a more detailed structure. However, this simple example is useful as an estimation for the range of accurate operation of our code. 

\subsection{Emitter-to-observer scheme}

For a convergence test of the emitter-to-observer scheme we chose the (modified) rigidly orbiting hot spot model described in Section~\ref{sec:hot_spot}. The radius of the spot is $R_{\mathrm{spot}}=0.25M$. We compared seven different runs sampling $N_i = 2 \times 2^i$ million photon packages ($0 \leq i \leq 6$). In Fig.~\ref{fig:convergence} we show the relative $L_2$-errors for the light curves with respect to the highest-resolution run, namely
\begin{equation}
    e_{i} = \frac{ || F_i(t) - F_6(t)||_{L_2}}{||F_6(t)||_{L_2}}\,,\quad 0 \leq i \leq 5\,.
\end{equation}
A least-squares linear fit of the error data gives the relation $\log_{10}(e_i) \approx -0.87 - 0.18 i$, meaning the error approximately follows a power law 
\begin{equation}
e_i \approx 10^{-2} \left( \frac{N_i}{\num{64e6}}\right)^{-0.6}\,
\end{equation}
in terms of the number of photon packages in the sample. Note that the index of the power law is consistent with what is commonly expected from Monte Carlo simulations.

\begin{figure}
    \includegraphics[width=\columnwidth]{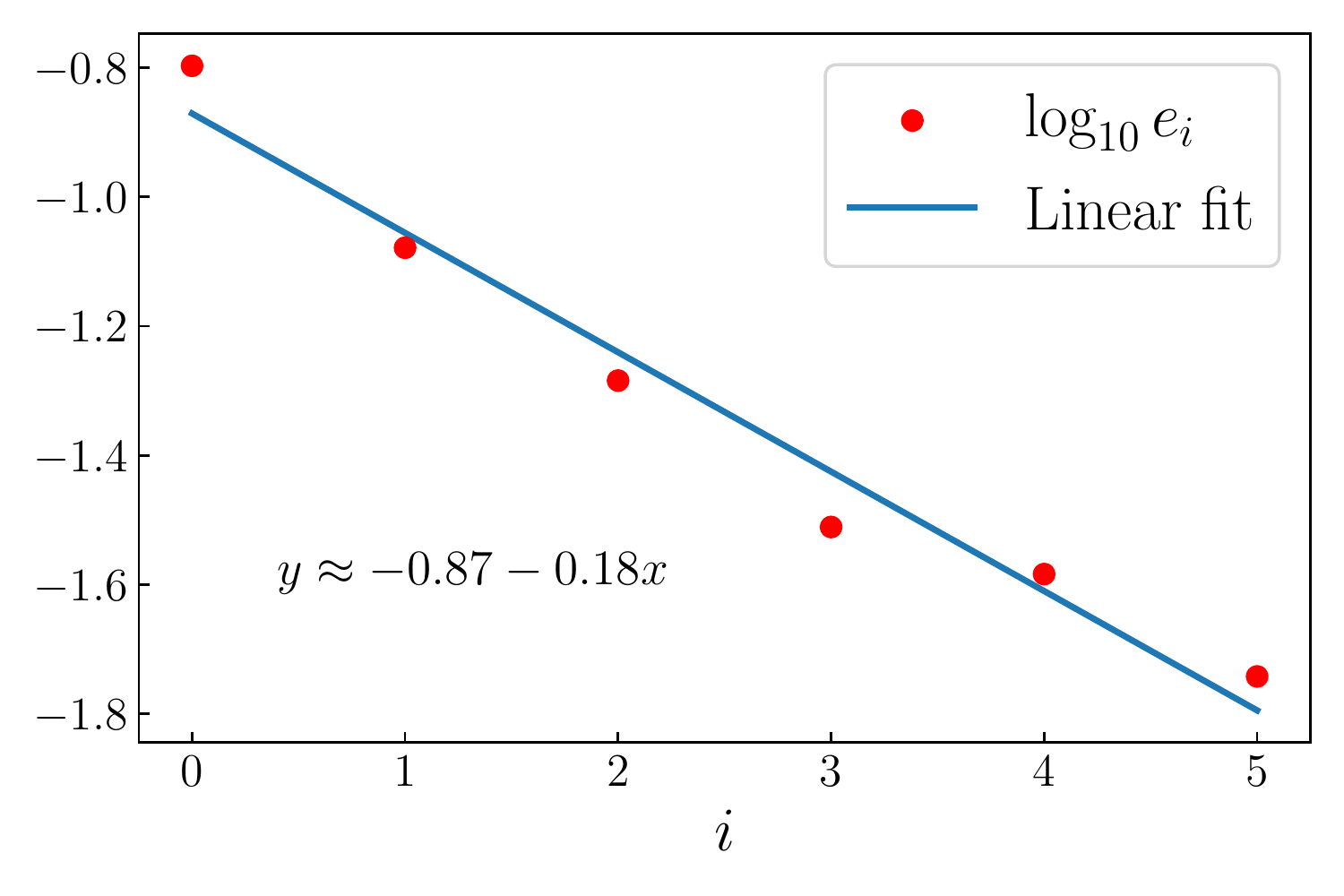}
    \caption{The dots correspond to the relative $L_2$-errors for the convergence test of the emitter-to-observer scheme in the orbiting hot spot model with a total of $N_i = 2 \times 2^i$ million photon packages. The solid line is the least squares linear fit of the data points. The error approximately follows a power law with index $p = -0.6$ in terms of the number of photon packages.}
    \label{fig:convergence_direct}
\end{figure}

\section{Conclusions}
\label{sec:conclusions}

We have presented \texttt{Skylight}, a new general-relativistic ray-tracing and radiative transfer code for calculating the observable quantities associated to astrophysical models of compact objects. One of the strengths of \texttt{Skylight} is its flexibility, in that it supports arbitrary space-time geometries and coordinate systems. We have kept this flexibility because in the near future we will apply the code to problems which do not restrict to the Kerr metric, particularly with approximate and numerical metrics of compact binary systems. The code has two equivalent schemes of operation, an observer-to-emitter scheme and an emitter-to-observer scheme, and it is capable of producing images, spectra and light curves. 

We have verified the correctness of our ray-tracing algorithm by checking the constants of motion and comparing with a semianalytic ray-tracing function, demonstrating the great accuracy of the integrator. Additionally, we have proved the usefulness of both operation schemes of \texttt{Skylight} for astrophysical applications by testing them in various problems and comparing with the literature: the relativistic broadening of an emission line from a thin accretion disk around a Kerr black hole, a hot spot orbiting a Schwarzschild black hole, and a hot spot over the surface of a spinning neutron star. This involves tests of the spectral and temporal dependencies of the code, and the management of units and normalization of curves. In all cases we obtained an excellent agreement with the results in the literature, having checked also the mutual equivalence of both \texttt{Skylight} schemes and the expected convergence rates.

Besides, this is the first Julia ray tracer which can handle arbitrary space-time geometries. This means our work also demonstrates the suitability of the relatively new language Julia for applications to highly-demanding computing problems in astrophysics.

As mentioned before, our plan is to explore diverse electromagnetic emission models and apply \texttt{Skylight} to various astrophysical scenarios, starting from emission models built on top of the corresponding force-free numerical solutions. Currently, we are already working on the X-ray light curves of millisecond pulsars from the FF pulsar solutions in \cite{carrasco2018pulsar} (which take into account the space-time curvature) and based on the simple emission model proposed in \cite{lockhart2019x}. 
In the near future, we plan to use the code to search for observable features on magnetar X-ray light curves associated to outburst events (like those studied in \cite{carrasco2019triggering}). And, further, we also aim at adapting \texttt{Skylight} to investigate the inspiral phase of black hole-neutron star binary systems, considering some candidate emission models based on the numerical solutions of e.g. \cite{carrasco2021magnetospheres} to probe the signatures of potential precursor electromagnetic signals in different bands of the spectrum.

\section*{Acknowledgements}

We would like to thank Dr. Christopher Rackauckas, lead developer of the package DifferentialEquations.jl that we use, for his valuable help with parallelization and the choice of solver methods.

We acknowledge financial support from CONICET, SeCyT-UNC, and MinCyT-Argentina.
Numerical computations were performed on the Sakura cluster at Max-Planck Computing and Data Facility, and on the Serafin Cluster (https://ccad.unc.edu.ar/equipamiento/cluster-serafin/) at Centro de Computación de Alto Desempeño, Universidad Nacional de Cordoba, which is part of the Sistema Nacional de Computación de Alto Desempeño, MinCyT-Argentina.

\section*{Data Availability}

No new data were generated or analysed in support of this research.



\bibliographystyle{mnras}
\bibliography{Skylight} 




\appendix

\section{Christoffel symbols of the Kerr metric in Kerr-Schild Cartesian coordinates}
\label{app:appendix}

Since all the components of the Kerr metric in Kerr-Schild coordinates are nonzero, obtaining the Christoffel symbols analytically to use them in the ray tracer implies quite involved calculations. To the best of our knowledge, they have not been explicitly written in the literature before, so we give some details here for future reference.

A general Kerr-Schild metric has the form
\begin{equation}
\label{eq:kerrmetric_app}
g_{\mu\nu}= \eta_{\mu\nu}+ 2 H l_\mu l_\nu\,,
\end{equation}
where $\eta_{\mu\nu}$ is the flat metric, $H$ is a scalar function and $l_\mu$ is a null covector. From equation~(2.3) of \citet{gurses1975lorentz}, we can derive the following expression for the Christoffel symbols of a general metric in Kerr-Schild form:
\begin{equation}
    \begin{aligned}
    \Gamma^\alpha_{\mu \nu} &=  \partial_\mu(Hl_\nu l^\alpha) + \partial_\nu(Hl_\mu l^\alpha) - \partial^\alpha(Hl_\mu l_\nu)\, + \\
    &+ 2Hl^\alpha l_\mu l_\nu l^\rho \partial_\rho H\,.
\end{aligned}
\label{eq:christoffel}
\end{equation}
Thus, we only need to compute the derivatives of $H$ and $l_\mu$ and insert them into equation~(\ref{eq:christoffel}). In the special case of the Kerr metric, these derivatives are:
\begin{equation}
    \begin{aligned}
    \partial_x r &= \frac{x r^3(r^2+a^2)}{a^2z^2(2r^2+a^2)+r^4(x^2+y^2+z^2)}\,, \\
    \partial_y r &= \frac{y r^3(r^2+a^2)}{a^2z^2(2r^2+a^2)+r^4(x^2+y^2+z^2)}\,, \\
    \partial_z r &= \frac{z r(r^2+a^2)^2}{a^2z^2(2r^2+a^2)+r^4(x^2+y^2+z^2)}\,.
    \end{aligned}
\label{eq:deriv_r}
\end{equation}
These quantities satisfy the useful relation $l_x \partial_x r + l_y \partial_y r + l_z \partial_z r = 1$. On the other hand, we have
\begin{equation}
    \begin{aligned}
    \partial_x H &=  \frac{-M(r^6 - 3 a^2 r^2 z^2)} {(r^4+a^2 z^2)^2} \partial_x r\,, \\
    \partial_y H &=  \frac{-M(r^6 - 3 a^2 r^2 z^2)} {(r^4+a^2 z^2)^2} \partial_y r\,, \\
    \partial_z H &= \frac{-Mr^2 [2a^2 r z + (r^4-3a^2z^2)\partial_z r ]} {(r^4+a^2 z^2)^2}\,,\\
    l^\rho \partial_\rho H &=  \frac{-Mr^2(r^4-a^2z^2)} {(r^4+a^2 z^2)^2}\,,
    \end{aligned}
\label{eq:deriv_H}
\end{equation}
and, finally,
\begin{equation}
    \begin{aligned}
    \partial_x l_x &= \frac{(x-2r l_x)\partial_x r+r} {r^2+a^2}  \,, \\
    \partial_y l_x &= \frac{(x-2r l_x)\partial_y r+a} {r^2+a^2} \,, \\
    \partial_z l_x &= \frac{(x-2r l_x)\partial_z r} {r^2+a^2}  \,,\\
    \partial_x l_y &= \frac{ (y-2r l_y)\partial_x r-a} {r^2+a^2} \,, \\
    \partial_y l_y &= \frac{ (y-2r l_y)\partial_y r+r} {r^2+a^2} \,, \\
    \partial_z l_y &= \frac{ (y-2r l_y)\partial_z r} {r^2+a^2}  \,,\\
    \partial_x l_z &= -\frac{z}{r^2}\partial_x r \,, \\
    \partial_y l_z &= -\frac{z}{r^2}\partial_y r \,, \\
    \partial_z l_z &= \frac{1}{r}-\frac{z}{r^2}\partial_z r \,.
    \end{aligned}
\label{eq:deriv_l}
\end{equation}
The expressions are cumbersome, but we only need to introduce them into the code and add them together appropriately, according to equation~(\ref{eq:christoffel}).


\end{document}